# Observation of plaid-like spin splitting in a noncoplanar antiferromagnet

Yu-Peng Zhu[1,9], Xiaobing Chen[1,9], Xiang-Rui Liu[1], Yuntian Liu[1], Pengfei Liu[1], Heming Zha[2], Gexing Qu[3], Caiyun Hong[4], Jiayu Li[1], Zhicheng Jiang[2], Xiao-Ming Ma[1], Yu-Jie Hao[1], Ming-Yuan Zhu[1], Wenjing Liu[2], Meng Zeng[1], Sreehari Jayaram[5], Malik Lenger[5], Jianyang Ding[2], Shu Mo[1], Kiyohisa Tanaka[6], Masashi Arita[7], Zhengtai Liu[2], Mao Ye[2], Dawei Shen[2], Jörg Wrachtrup[5], Yaobo Huang[8], Rui-Hua He[4], Shan Qiao[2†], Qihang Liu[1†], Chang Liu[1†]

[1]Shenzhen Institute for Quantum Science and Engineering (SIQSE) and Department of Physics, Southern University of Science and Technology (SUSTech), Shenzhen, Guangdong 518055, China

[2]National Key Laboratory of Materials for Integrated Circuits, Shanghai Institute of Microsystem and Information Technology, Chinese Academy of Sciences, Shanghai 200050, China

[3]Beijing National Laboratory for Condensed Matter Physics and Institute of Physics, Chinese Academy of Sciences, Beijing 100190, China

[4]Key Laboratory for Quantum Materials of Zhejiang Province, Department of Physics, Westlake University, Hangzhou, Zhejiang 310024, China

[5]3rd Institute of Physics, IQST and Centre for Applied Quantum Technologies, University of Stuttgart, Stuttgart 70569, Germany

[6]National Institutes of Natural Science, Institute for Molecular Science, Okazaki 444-8585, Japan

[7]Hiroshima Synchrotron Radiation Center, Hiroshima University, Higashi-Hiroshima 739-0046, Japan

[8]Shanghai Synchrotron Radiation Facility, Shanghai Advanced Research Institute, Chinese Academy of Sciences, Shanghai 201204, China

[9]These authors contributed equally: Yu-Peng Zhu, Xiaobing Chen.

†e-mail: qiaoshan@mail.sim.ac.cn, liuqh@sustech.edu.cn, liuc@sustech.edu.cn

Spatial, momentum and energy separation of electronic spins in condensed matter systems guides the development of novel devices where spin-polarized current is generated and manipulated[1-3]. Recent attention on a set of previously overlooked symmetry operations in magnetic materials[4] leads to the emergence of a new type of spin splitting, enabling giant and momentum-dependent spin polarization of energy bands on selected antiferromagnets[5-10]. Despite the ever-growing theoretical predictions, the direct spectroscopic proof of such spin splitting is still lacking. Here, we provide solid spectroscopic and computational evidence for the existence of such materials. In the noncoplanar antiferromagnet $MnTe_2$, the in-plane components of spin are found to be antisymmetric about the high-symmetry planes of the Brillouin zone, comprising a plaid-like spin texture in the antiferromagnetic (AFM) ground state. Such an unconventional spin pattern, further found to diminish at the high-temperature paramagnetic state, stems from the intrinsic AFM order instead of spin-orbit coupling (SOC). Our finding demonstrates a new type of quadratic spin texture induced by time-reversal breaking, placing AFM spintronics on a firm basis and paving the way for studying exotic quantum phenomena in related materials.

Historically, mechanisms for splitting a spin-degenerate energy band (Fig. 1a) include Zeeman interaction[11] in ferromagnets where electrons with different spins are equally separated in energy regardless of their momenta (Fig. 1b), and Rashba/Dresselhaus interaction[12,13] in which the spins split in a momentum-dependent manner in nonmagnetic crystals lacking inversion symmetry (Fig. 1d). Recently, a novel type of spin splitting induced by the long-range magnetic order has been predicted in certain antiferromagnets, even when SOC is absent (Fig. 1c). Such proposal enables the choice of light-element materials for generation of e.g. spin currents[14-17] and tunnelling magnetoresistance effect[18-22], significantly widening the scope of AFM spintronics[23,24].

Instead of the well-studied (magnetic) space groups, the symmetry operations allowed in these novel magnets are fully described by an enhanced symmetry group called spin space group[4,25-31], where the decoupling of lattice rotation and spin rotation permits certain symmetry operations that are excluded by conventional magnetic space groups, leading to exotic physical phenomena such as weak-SOC $Z_2$ topological phases[4], chiral Dirac-like fermions[32,33], *C*-pair spin valley locking[34], and anomalous Hall effect[7,35,36]. The appearance of AFM-induced spin splitting is unique in that its splitting scale depends on the momentum of the electrons and can be much larger than the largest known Rashba effect[37]. The collinear members of these antiferromagnets are named "altermagnets" in the literature[9,38], while such unconventional spin splitting can also take place in noncollinear magnets[6].

Despite the thriving theoretical studies on these novel magnets[4,6-10] and transport evidence on one of them[15-17,35], the direct spectroscopic proof of such spin splitting itself is lacking. Here, using spin- and angle-resolved photoemission spectroscopy (SARPES) measurements and theoretical analysis, we demonstrate unambiguously the existence of this AFM-induced spin splitting effect on a noncoplanar antiferromagnet, manganese ditelluride ($MnTe_2$). By virtue of the state-of-the-art SARPES facilities[39,40], we observe that the in-plane components of the spin on a bulk band are antisymmetric about both horizontal and vertical high-symmetry planes. Thus, the spin texture forms a plaid pattern in the three-dimensional Brillouin zone, consistent with our calculations. Temperature-dependent SARPES measurements further indicate that the observed spin polarization is related to the antiferromagnetic-paramagnetic transition. Our work uncovers a quadratic spin splitting effect induced by magnetic exchange, distinct from the well-known Zeeman, Rashba and Dresselhaus scenarios, in terms of the spin texture as well as the underlying mechanism (Extended Data Table 1).

## Calculated spin polarization in MnTe$_2$

We begin with the essential physical properties of the MnTe$_2$ single crystals (see Methods and Extended Data Fig. 1 for details). It is interesting and rare that the crystalline space group, the magnetic space group, and the spin space group of MnTe$_2$, shown in Fig. 1e, are the same (*Pa-3*), indicating that the noncoplanar AFM order and SOC do not break any spatial symmetries. Because of these groups, MnTe$_2$ has neither *PT* (spatial inversion followed by time reversal) nor $U\tau_{1/2}$ (unitary spin inversion followed by fractional translation) symmetry, satisfying the design principle of the momentum-dependent spin splitting purely induced by the AFM order[4,6].

Figure 1f shows the first Brillouin zone of MnTe$_2$, in which we mark the high symmetry plane Γ-X1-M-X and the ARPES-measured, parallel O-A-B-C plane with $k_z = -0.2\ \pi/c$ (see Extended Data Fig. 2 for the discussion of $k_z$ dispersion). Figure 1g-i are schematics of the $S_x$, $S_y$, and $S_z$ components of the valence band, where the blue and red plaids represent the positive and negative spin polarization component, respectively. The spin-resolved *E-k* dispersion and constant energy contours (CECs) through density functional theory (DFT) calculation, in Fig. 1j-l, reveal two main features. First, instead of SOC, the spin splitting of the itinerant electron is determined by the inhomogeneous magnetic field from the AFM order. The DFT-calculated *E-k* dispersion without SOC (Fig. 1j) exhibits a large spin splitting for the O-A-B-C plane, with comparable magnitude to the SOC-included case (Extended Data Fig. 3a), even though tellurium is expected to have a pronounced SOC effect on the electronic structure. Second, a unique plaid-like spin texture with alternating polarizations on each side of a high symmetry plane is predicted (Fig. 1g-i, l). This is because the four face-centered cubic Mn sublattices with different spin orientations are connected by three magnetic group mirror symmetries $M_x$, $M_y$, and $M_z$. The spin component $S_x$ is antisymmetric about the $k_y = 0$ plane and the $k_z = 0$ plane but symmetric about the $k_x = 0$ plane due to $M_y$ ($S_x \Rightarrow -S_x$, $S_y \Rightarrow S_y$, $S_z \Rightarrow -S_z$) and $M_z$ ($S_x \Rightarrow -S_x$, $S_y \Rightarrow -S_y$, $S_z \Rightarrow S_z$), while the other two components $S_y$ and $S_z$ have analogous transformations. The high-symmetry Γ-X1-M-X plane ($k_z = 0$) manifest no in-plane spin splitting because $M_z$ forbids in-plane spin components (Extended Data Table 2). Therefore, we choose the O-A-B-C plane for the SARPES measurements.

*Spin polarization at a fixed $k_z$*

We next investigate the spin-polarized bands of MnTe$_2$ with systematic SARPES measurements in its AFM phase ($T$ = 30 K). In the dataset of Fig. 2, photoelectrons for both spin-resolved and spin-integrated measurements were excited by a He lamp with $h\nu$ = 21.2 eV. This photon energy is found to correspond to $k_z$ = –0.2 $\pi/c$, i.e. the O-A-B-C plane in Fig. 1f. The bands observed by ARPES are bulk electronic states with substantial $k_z$ broadening (see Methods, Extended Data Fig. 2 and 4 for more discussion). As illustrated in Fig. 2b, the measured CEC has the shape of a cross with arms along the O-A and O-C directions, consistent with the DFT-calculated bulk bands (Fig. 2c, d). It is clear that the DFT bulk bands are highly spin-polarized, with in-plane polarization vectors $S_x$ ($S_y$) antisymmetric about the O-A (O-C) line. Such unique AFM-induced, plaid-like spin polarization is confirmed by our SARPES measurements. To see this, we present the $S_x$- and $S_y$-resolved $E$-$k$ images along Cuts 1 – 4 labeled in Fig. 2b. Each image is about ±0.22 Å$^{-1}$ (i.e., ±0.5 $\pi/a$) away from O-A and O-C, and thus expected to have antisymmetric in-plane spin polarization.

Figure 2e-j present spin-integrated-, DFT calculated-, and spin-resolved ARPES band dispersion, as well as their spin-resolved energy distribution curves (EDCs), and the curves of spectral intensity and spin polarization versus binding energy and momentum, respectively. The full set of raw data for spin-resolved band dispersion is presented in Extended Data Fig. 5. Overall, a nice consistency between the DFT-extracted spin-polarization and that measured by SARPES is reached. A closer look into the SARPES data leads to several interesting observations. First, the $S_x$-resolved $E$-$k$ maps along Cut 1 and Cut 2 (shown in Fig. 2e, f) exhibit opposite $S_x$-polarization, which is caused by the mirror reflection $M_y$: $S_x \Rightarrow -S_x$. In addition, the $S_x$-polarization in Cut 1 and Cut 2 are opposite at the top and shoulders of VB1, which is also well reproduced by DFT calculations with and without SOC (Extended Data Fig. 3). Second, for Cuts 3 and 4, the $S_x$-polarization of VB1 is antisymmetric about $k_y$ = 0 (Fig. 2g, h). Third, the situation for the spin component $S_y$ is opposite. The measured spectra along Cut 1 and Cut 4 (Fig. 2i, j) exhibit a high level of $S_y$-polarization that are antisymmetric about O-C ($k_x$ = 0) due to the mirror reflection $M_x$: $S_y \Rightarrow -S_y$,

consistent with the DFT calculations. The same conclusions can also be drawn with the spin EDCs and polarization curves. Furthermore, the SARPES data along Cuts 5 and 6 shown in Extended Data Fig. 6 demonstrates that the in-plane spin texture is Dresselhaus-like, with substantial radial components along the diagonal directions of the in-plane Brillouin zone. Based on such data, we conclude that momentum-dependent antisymmetric spin splitting with plaid-like alternating spin texture exists in the noncoplanar antiferromagnet $MnTe_2$ in its magnetic ground state.

*Antisymmetric spin components along $k_z$*

After establishing the existence of the spin-polarized bands in the AFM phase of $MnTe_2$, a crucial topic is to experimentally differentiate the observed spin texture from those induced by SOC in the surface states. The characteristic sign reversal of the spin at opposite out-of-plane momenta ($k_z$'s) would pinpoint the AFM origin of the observed polarization, as no $k_z$ dispersive behavior of the spin is expected in any spin splitting effects originated from surface inversion breaking. In Fig. 3 we show such $k_z$-dependent sign change. We present SARPES data taken at a synchrotron-based SARPES setup at four different incident photon energies, corresponding to four different $k_z$'s: 21.2 eV ($k_z = -0.2\ \pi/c$), 28 eV ($k_z = 0.5\ \pi/c$), 66 eV ($k_z = -0.5\ \pi/c$), and 82 eV ($k_z = 0.5\ \pi/c$). At each photon energy, polarization curves are measured at one or two $k$ points, as marked "L" and "R" in Figure 3c. According to DFT calculations on the bulk bands (Fig. 3a, b), the sign of $S_x$ will reverse across the high symmetry planes $k_z = 0$ and $k_y = 0$. Indeed, this is what we observed experimentally in Fig. 3c. At the "L" position, $S_x$ is found to reverse the sign between 21.2 eV ($S_x > 0$) and 28 eV ($S_x < 0$), and between 66 eV ($S_x > 0$) and 82 eV ($S_x < 0$); at the "R" position, $S_x$ is found to reverse the sign between 21.2 eV ($S_x < 0$) and 28 eV ($S_x > 0$). Across the in-plane high symmetry plane of $k_y = 0$, $S_x$ is also found to reverse the sign between the "L" and "R" points at both 21.2 eV and 28 eV, further verifying the antisymmetric plaid-like feature within a certain $k_z$ plane.

To eliminate the effect of possible matrix element associated with different light polarizations, we measured the $S_x$ polarization at the "L" point at $h\nu = 28$ eV under two different incident beam polarizations, LH and LV, which adds up to an unpolarized situation (see Methods for more discussion). Both curves give $S_x < 0$ for most part of the binding energies. Therefore, we argue that the measured polarization reflects the intrinsic spin of the band. The $k_z$-dependent spin polarization is summarized in Fig. 3d. We mark the measured sign of $S_x$ polarizations (color of the dots) within a schematic $k_y$-$k_z$ frame, in which the plaid-like background represents the calculated sign of $S_x$. The nice consistency consolidates that the plaid-like spin texture is observed also along the out-of-plane momentum direction.

*Temperature dependence*

To verify the magnetic origin of such spin splitting, another critical piece of evidence is the temperature evolution of spin polarization[41]. Temperature changes are not expected to influence the strength of relativistic spin-orbit coupling, which is primarily dependent on atomic mass. In contrast, for AFM-induced spin splitting, the spin polarization is expected to disappear when the system evolves to the paramagnetic state above the Néel temperature $T_N$ (87 K). Our data in Fig. 4 supports the latter case. Figure 4a, b shows the spin-integrated band dispersion along Cut 1 at 30 and 110 K, as well as along Cut 3 at 30 and 150 K. From both the raw data and the second derivative result, we notice that the overall band structure undergoes significant modification as temperature rises above $T_N$ = 87 K (see Extended Data Fig. 7 and 8 for more details). Importantly, such change of bands in the high-$T$ paramagnetic phase is accompanied with an abrupt decrease of spin polarization. In Figure 4c and d, we show the temperature-dependent $S_x$ polarization curves along Cut 1 and Cut 3. It is apparent that signatures of $S_x$ polarization almost vanishes at high temperatures, in sharp contrast to the curves at 30 K where sign-changing $S_x$ components are seen for different binding energies / in-plane momenta.

Additional evidence is presented in Extended Data Fig. 9 where we plot the evolution of spin splitting along Cut 5 across $T_N$, as well as spin-resolved ARPES $E$-$k$ images along Cut 1 at 30 and 60 K. The pair of bands along Cut 5 seemingly split in momentum like the Rashba/Dresselhaus pairs at 30 K is found to merge into a single band at 150 K (Extended Data Fig. 9a, b), evident for the vanishing of spin splitting above $T_N$. The same data also gives an estimation of the spin splitting energy level: about 274 ± 40 meV at $k_\parallel$ = ±0.18 Å$^{-1}$. This value is comparable to the largest-known "giant Rashba splitting" found in BiTeI[37]. In the SARPES images (Extended Data Fig. 9c, d), the blue ($S_x > 0$) part at 30 K becomes mix-colored at 60 K, indicative of a decrease in spin polarization, while the overall shape and $E$-$k$ location of the band itself is unchanged. Taken collectively, our data likely identify that the observed spin-splitting effect in MnTe$_2$ is induced by the AFM order.

*Quadratic spin texture*

Our experimental results - the sign reversal of spin across $k_z = 0$ (Fig. 3) and the decrease of polarization across $T_N$ (Fig. 4) - serve as evidences for an AFM-induced spin splitting in MnTe$_2$. We now discuss more deeply the uniqueness of the observed plaid-like spin texture compared with the classical spin textures caused by SOC. For simplicity, we consider systems with inversion symmetry $P$ in the bulk, like MnTe$_2$, where bulk Dresselhaus and Weyl spin splitting are prohibited. Thus, the SOC-induced spin splitting is associated with $P$ breaking at the surface. Since SOC does not break time-reversal $T$, symmetry requirements enforce that the spin-splitting $k \cdot p$ Hamiltonian expanded at time-reversal-invariant momenta (TRIM) only allow odd-order $k$-polynomial, such as linear/cubic Rashba or Dresselhaus terms. In sharp contrast, when considering AFM order as the mechanism of spin splitting, the existence of $P$ and breaking of $T$ results in the $k \cdot p$ Hamiltonian at TRIM points having only even-order $k$-polynomial[42]. In the case of MnTe$_2$, the Hamiltonian expanded near the Γ point is written as $k_z(k_y\sigma_x + k_x\sigma_y)$ (see Methods for details). Such a quadratic spin texture naturally leads to the sign reversal of spin across $k_z = 0$.

Extended Data Table 1 summarizes the critical symmetries, observable characters, the nature of $k \cdot p$ Hamiltonians and spin patterns of the two distinct spin splitting phenomena. We note that although at $k_z > 0$, plaid-like and linear Dresselhaus spin splitting share qualitatively the same in-plane spin patterns, the $k_z$ dependence of the former unveils the nature of an unprecedentedly measured quadratic spin texture, which is underlaid by the mechanism of time-reversal breaking.

In summary, our systematic SARPES measurements have demonstrated the existence of a new type of momentum-dependent spin splitting induced by the intrinsic AFM order in a noncoplanar antiferromagnet MnTe$_2$. We emphasize that this type of spin splitting in noncoplanar and noncollinear antiferromagnets has the same origin as that in collinear altermagnets, as the local AFM field couples the electron spin and its motion in the same manner. The momentum-dependent spin-splitting bands in MnTe$_2$ can efficiently generate spin-polarized current, leading to magnetic spin Hall effec[43-45], spin-splitter effect, and tunnelling magnetoresistance, etc. Furthermore, such spin-splitting effect could also exist in a variety of quantum materials such as Mott insulators[46], parent compounds of unconventional superconductors[47], and three-dimensional quantum Hall materials[48,49], providing an avenue to study these exotic phases of matter and potential applications in spintronics.

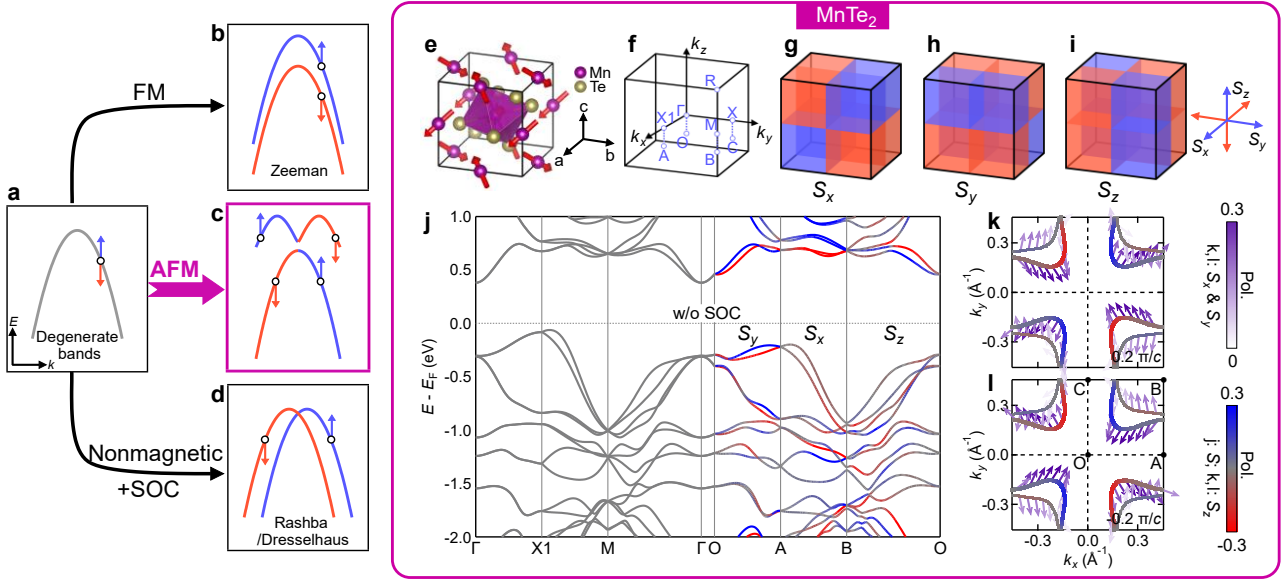

**Fig. 1 | Different prototypes of spin-splitting effect, and DFT calculation results of MnTe$_2$. a-d**, Schematics of the spin degenerate bands and spin splitting induced by the Zeeman effect, AFM order, and spin-orbit interaction with structural asymmetry. Blue and red denote opposite spins. **e**, Crystal structure of MnTe$_2$. The noncoplanar magnetic configuration of Mn atoms is indicated by the red arrows. Magnetic moments of about 4.3 $\mu_B$ are arranged along the bulk diagonal direction. **f**, The first Brillouin zone of MnTe$_2$. The O-A-B-C plane is the $k_z = -0.2$ $\pi/c$ plane, corresponding to the plane of our ARPES and SARPES measurements with 21.2 eV photons. **g-i**, Schematics for the sign of the $S_x$, $S_y$ and $S_z$ polarization in the three dimensional Brillouin zone, showing plaid-like antisymmetric spin texture across the high symmetry planes. **j**, DFT-calculated spin-resolved bands without SOC. There is no in-plane spin polarization in the $k_z = 0$ (Γ-X-M-X1) plane, while a high level of spin polarization is seen in the O-A-B-C plane. **k,l**, DFT-calculated spin-resolved CEC at $k_z = \pm 0.2$ $\pi/c$ and at binding energy $E_B = 0.45$ eV. Magenta arrows: in-plane direction of the spin; darkness of arrows: magnitude of in-plane spin polarization; color of bands: magnitude of the out-of-plane $S_z$ polarization. Clearly, $S_\alpha$ is antisymmetric about the $k_\beta = 0$ and $k_\gamma = 0$ planes, and symmetric about the $k_\alpha = 0$ plane ($\alpha$, $\beta$, $\gamma = x, y, z$).

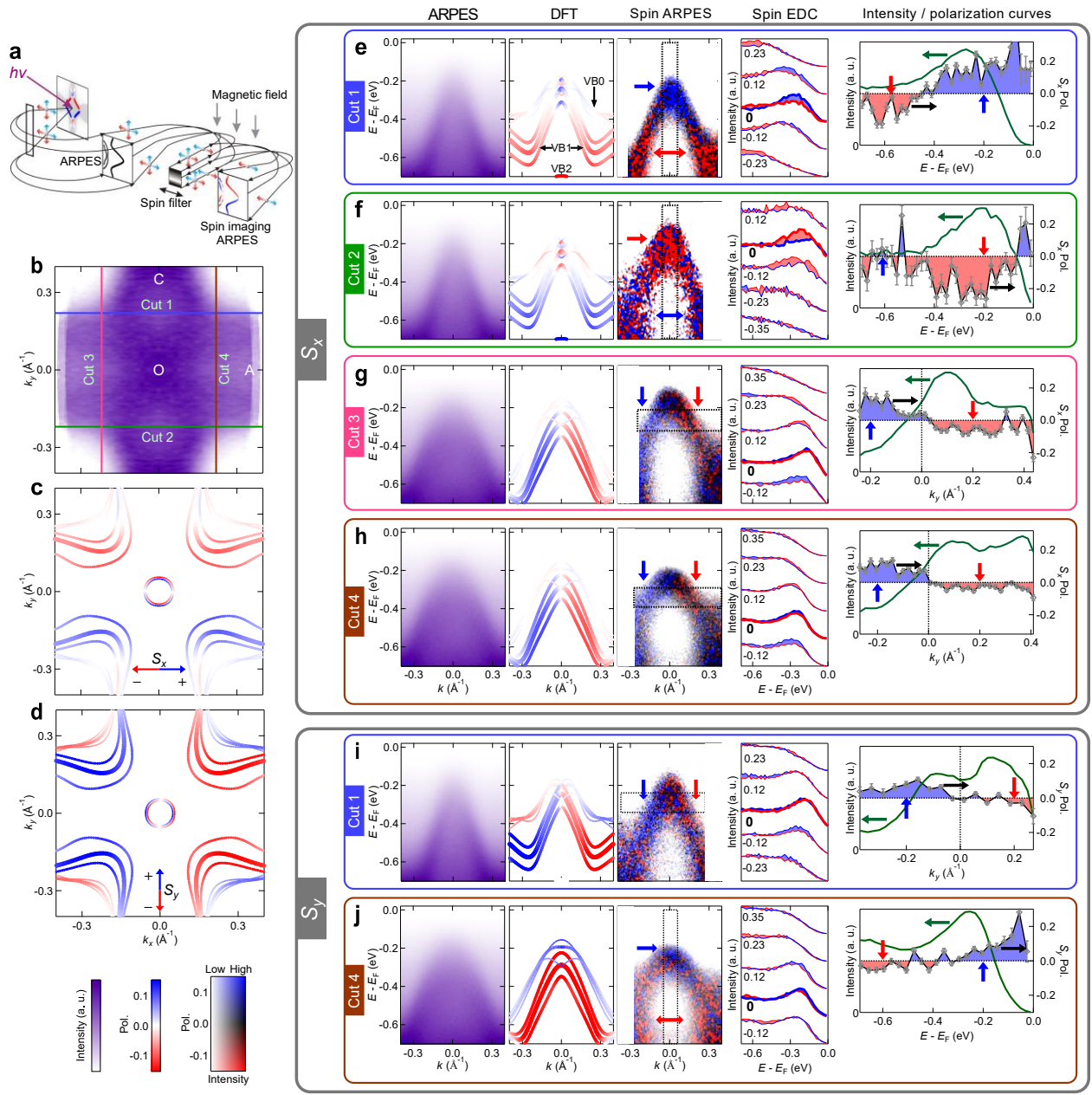

**Fig. 2 | Plaid-like texture of in-plane spins at $k_z = -0.2\ \pi/c$. a**, Schematic setup of the spin-imaging ARPES system (see Methods). **b**, ARPES-derived, spin-integrated CEC at binding energy $E_B = 0.45$ eV. **c,d**, Corresponding DFT-calculated, $S_x$- and $S_y$-resolved CECs of bulk electronic states. The $k_z$-broadening effect is considered by plotting the calculated CECs at $k_z = -0.1, -0.2$ (widened), and $-0.3\ \pi/c$ together. Blue (red) color represents the level of $S_x+$ ($S_x-$) / $S_y+$ ($S_y-$) polarization (same below). **e-j**, From left to right: spin-integrated ARPES band dispersion, DFT-calculated bulk bands, spin-resolved ARPES band dispersion, spin-resolved EDCs, and spin-integrated ARPES intensity (green curve) as well as spin polarization (symbols and filled black curve) versus binding energy or momentum. Spin-resolved ARPES band dispersion is multiplied by the total intensity ($I_{up} + I_{down}$) for better visualization of spin polarization on the bands. Red and blue arrows highlight the signs of the spin. The $k$ positions of spin-resolved EDCs are marked near the curve. Each point on a curve is an integrated intensity over a $(E, k)$ range of (20 meV, 0.12 Å$^{-1}$). The spin-integrated ARPES intensity and spin polarization are integrated within the dashed rectangles in spin-resolved ARPES panels. Error bars are defined as the standard deviation of polarization within the integrated region, which reflect the deviation of polarization from the mean value.

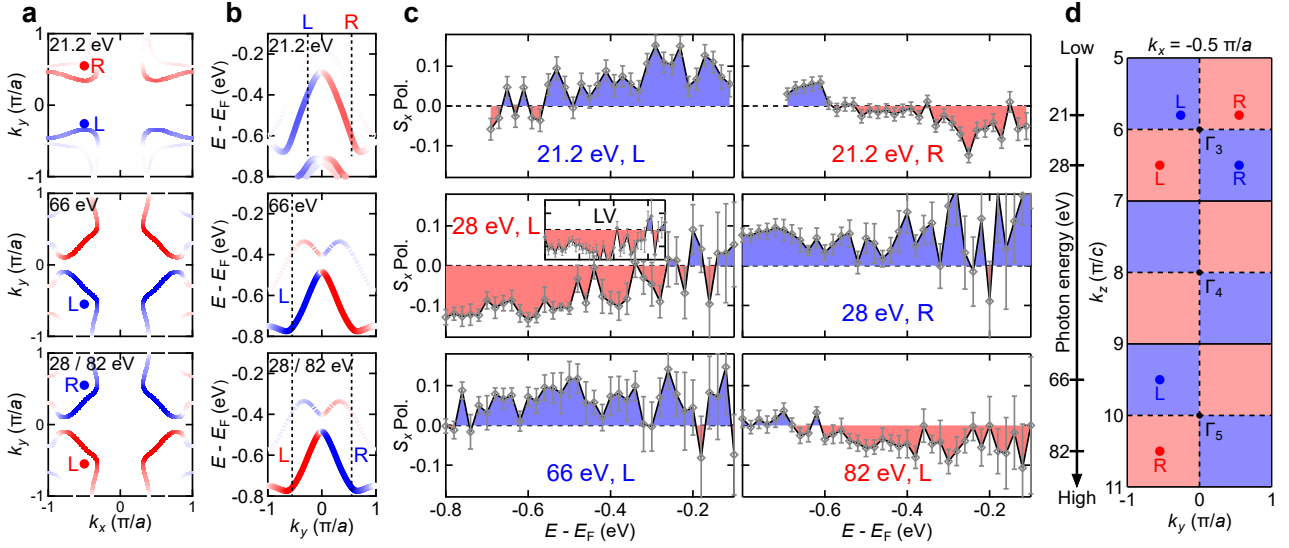

**Fig. 3 | Sign reversal of the $S_x$ polarization at different $k_z$'s. a**, DFT calculated, $S_x$-resolved CECs at $k_z = -0.2\ \pi/c$ (corresponding to 21.2 eV incident photons, at binding energy $E_B = 0.45$ eV), $-0.5\ \pi/c$ (66 eV, $E_B = 0.65$ eV), and $0.5\ \pi/c$ (28 and 82 eV, $E_B = 0.65$ eV). The "L" and "R" dots mark the in-plane momenta of the polarization curves shown in **c**. **b**, DFT-derived $S_x$-resolved $E$-$k$ dispersion at $k_x = -0.5\ \pi/a$, crossing the "L" and "R" points (dashed lines). $S_x$ is antisymmetric about the $k_y = 0$ and $k_z = 0$ planes due to the mirror symmetries $M_y$ and $M_z$. **c**, Measured $S_x$ polarization at the "L" and "R" points, at $k_z = -0.2\ \pi/c$ (21.2 eV), $0.5\ \pi/c$ (28 eV), $-0.5\ \pi/c$ (66 eV), and $0.5\ \pi/c$ (82 eV). Incident light is linearly polarized horizontally (LH), except for the inset of panel $0.5\ \pi/c$ (28 eV, L), where it is linearly polarized vertically (LV). Error bars are defined the same way as in Fig. 2. $S_x$ is found to be antisymmetric about both the $k_y = 0$ and the $k_z = 0$ plane, consistent with DFT calculations. **d**, Schematics of $S_x$ versus $k_z$ (i.e. versus photon energy), showing the characteristic plaid-like texture.

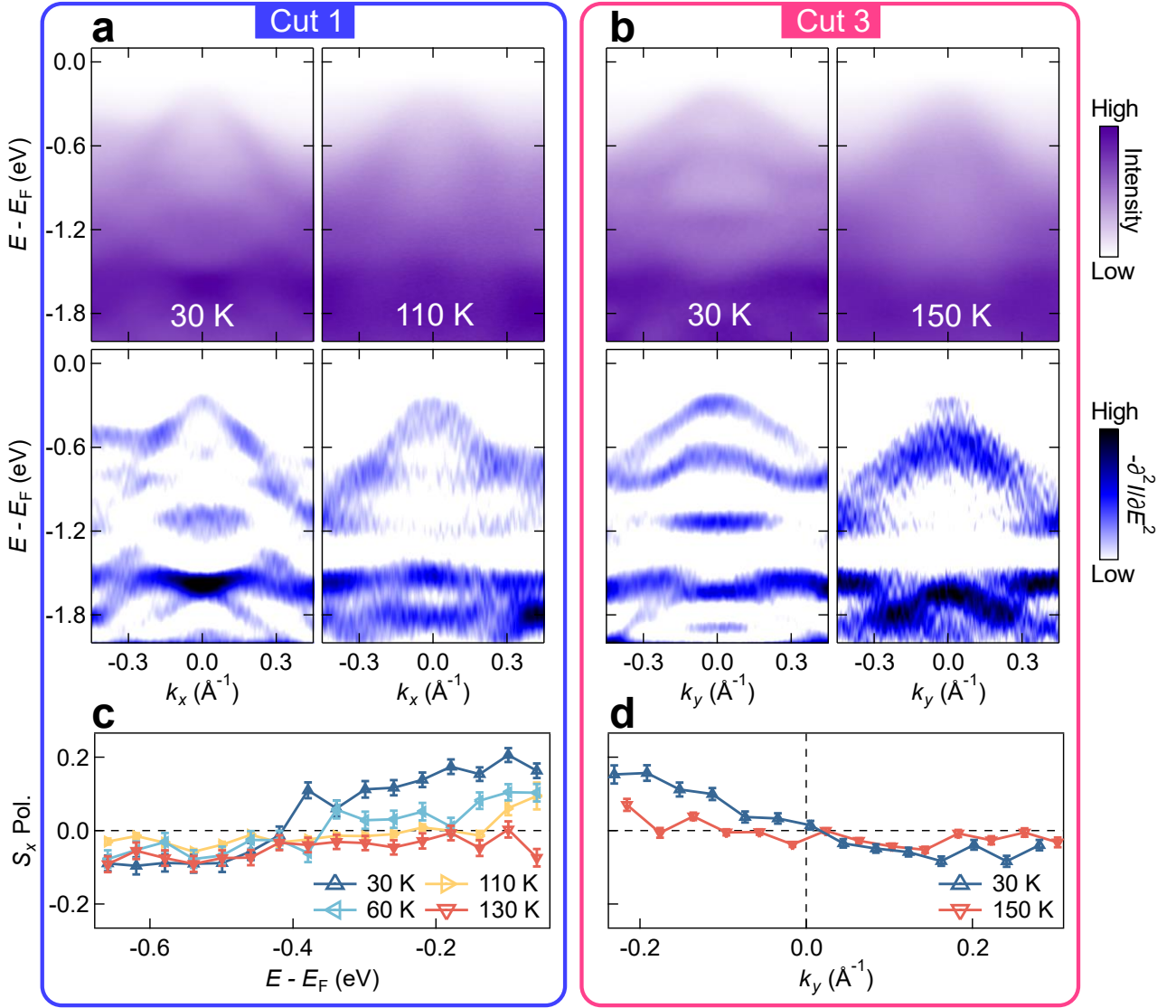

**Fig. 4 | Temperature dependence of the band structure and the $S_x$ polarization. a,b,** Spin-integrated band dispersion along Cut 1 and Cut 3 below and above $T_N = 87$ K, and corresponding second derivative results along the EDCs. **c,d,** Temperature evolution of $S_x$. The measurement positions are the same as Fig. 2e and g, respectively. Compared with the 30 K data, $S_x$ decreases by about one half at $T = 60$ K, and nearly becomes zero for $T > T_N$. This result supports the AFM-induced spin polarization mechanism, rather than the SOC-originated ones which are constant at all temperatures.

## Methods

### Sample growth and characterization

Single crystal of $MnTe_2$ were synthesized by the chemical vapor transport method. Manganese powder, tellurium powder, and iodine crystals were mixed and sealed into a silica tube under vacuum. After heating at 600 °C with a temperature gradient of 60 °C for 10 days, millimeter-size plate-shaped $MnTe_2$ single crystals were obtained, shown in Extended Data Fig. 1a.

$MnTe_2$ samples were characterized by X-ray diffraction using a Rigaku Smartlab diffractometer with Cu K$\alpha$ radiation at room temperature. The diffraction pattern in Extended Data Fig. 1a confirms the pyrite structure in the space group *Pa-3*, where the cleaving planes are parallel to the (001) crystallographic orientation. In the lattice, manganese and tellurium atoms are located at the 4a and 8c Wyckoff positions, respectively. The crystal is composed of face-centered cubic Mn sublattices connected by $MnTe_6$ octahedra.

In Extended Data Fig. 1b, the magnetic properties were measured by a Quantum Design Physical Property Measurement System with a magnetic field up to 7 T. The temperature dependence of the magnetization curve indicates that the Néel temperature is about 87 K, consistent with previous studies[50,51]. Residual magnetization below 87 K in seen in our field-cooled (FC) *M-T* curve, while an abrupt downturn of magnetization at 87 K is seen in our zero-field-cooled (ZFC) *M-T* curve. This signifies an antiferromagnetic behavior at low temperatures. The magnetic field dependence of the magnetization curve is linear, indicated that no ferromagnetism is observed in our measurements.

Extended Data Fig. 1c, d presents a scanning electron microscopy (SEM) image with corresponding energy-dispersive X-ray spectroscopy (EDS) elemental maps, measured with a ZEISS Merlin SEM setup at the SUSTech Core Research Facilities. The uniform color of the maps indicates a homogeneous distribution of the Mn and Te elements. Figure 1d shows further that the atomic ratio of Mn:Te is close to 1:2 on all areas measured. Small but systematic deviation to the stoichiometric value implies slight but uniform Te deficiency in the crystal, leading to a refined chemical composition $MnTe_{2-\delta}$ with $\delta \sim 0.18$. Based on this data, we deduce that the crystals we used are of high quality and single phased.

**Lab-based spin-integrated ARPES measurements**

Lab-based spin-integrated ARPES measurements were performed at Westlake University with a Scienta-Omicron DA30 electron analyzer using a He lamp with $h\nu$ = 21.2 eV as light source. The MnTe$_2$ samples were cleaved *in-situ* and measured in an ultra-high vacuum with a base pressure better than $1.5\times10^{-10}$ mbar. Angular and energy resolutions were set to 0.1° and 5 meV, respectively.

In Fig. 2b, the measured CEC shows two different arms along the O-A and O-C directions, respectively. The arm along O-A is found to be slightly "wider" than that along O-C, since there is no $C_4$ symmetry in this system despite its cubic crystal structure. Such subtle difference not only guides us to align the samples and determine the spin directions during the experiment, but also infers that the magnetic domains in the MnTe$_2$ crystals has one preferred orientation over the others. To confirm this, we performed both scanning nitrogen-vacancy (NV) magnetometry measurements as well as SARPES experiments before and after a 90° rotation of the sample. Results from both techniques (Extended Data Fig. 10) are supportive for a domain size comparable to the incident beam spot, or a preferable orientation of magnetic moments within the region of the incident beam.

Spin-integrated ARPES band dispersion along Cuts 1 – 4 (Fig. 2e-j) exhibit a single hole-like band near $E_F$, consistent with the calculation results, that the main hole-like band (VB1) coexist with another band (VB0) at lower binding energies (see also Extended Data Fig. 3). Although this additional band is almost invisible in Fig. 2, its ARPES intensity and associated spin polarization is found present along Cuts 5 and 6, which are 45° rotated with respect to Cuts 1 – 4 (Extended Data Fig. 6). Considering its reduced ARPES and SARPES signal (probably due to matrix element effect), we lower the spin signal of VB0 by one half on every theoretical *E-k* cut throughout the paper.

**Synchrotron-based spin-integrated ARPES measurements**

The $k_z$ dispersion data and the spin-integrated electronic structure at different photon energies in Extended Data Fig. 2 and 4 were performed at BL03U[52] of the Shanghai Synchrotron Radiation Facility (SSRF) and BL09A of the Hiroshima Synchrotron Radiation Center (HiSOR). Data at BL03U of SSRF is measured with a Scienta-Omicron DA30 electron analyzer and *p*-polarized light with photon energies between 50 to 160 eV. The sample was cleaved *in-situ* with a base pressure of $9\times10^{-11}$ mbar. Data at BL09A of HiSOR is measured with a SPECS PHOBIOS 150 electron analyzer and *p*-polarized light with photon energies between 11 to 40 eV. The sample was cleaved *in-situ* with a base pressure of $1\times10^{-10}$ mbar. Measurement temperature is 30 K for both facilities.

In Extended Data Fig. 2 and 4, an inner potential of $V_0 = 10$ eV is determined by comparing the experimental results with the DFT calculation results. The observed bands are seen to exhibit clear $k_z$ dispersive behavior, which is not present in surface states. Moreover, we show in Extended Data Fig. 11 the DFT-calculated electronic structures from a semi-infinite slab model on two possible surface terminations, differentiating bulk bands from surface states. It is found that the surface bands contribute negligibly to the spectral weight in the *E-k* regions we measured, endorsing the bulk nature of the observed bands in Fig. 2. Slight deviation between DFT results and the experimental data can be attributed to the $k_z$ broadening effect of the bulk states[53], which is considered in all theoretical panels in Fig. 2 by simultaneously plotting the DFT bulk bands for neighboring $k_z$'s.

**Lab-based SARPES measurements and data analysis**

Lab-based spin-resolved electronic structures were obtained via a multichannel very low energy electron diffraction (VLEED) spin polarimeter attached to a Scienta R3000 hemispherical analyzer at Shanghai Institute of Microsystem and Information Technology (SIMIT), the schematics of which is shown in Fig. 2a. A multi-channel SARPES *E-k* cut is obtained by guiding the ARPES photoelectrons through a VLEED type spin filtering crystal which is pre-magnetized along the ±*x* (horizontal) or ±*y* (vertical) direction. The reflected electrons are then recorded on the CCD camera. $S_x$- ($S_y$)-resolved ARPES data is obtained by subtracting the $S_x$+ ($S_y$+)-magnetized *E-k* map with the $S_x$− ($S_y$−)-magnetized one[39,40]. Photoelectrons were excited by He I*α* line with $h\nu = 21.2$ eV. The sample was cleaved *in-situ* with a base pressure of $1\times10^{-10}$ mbar. Angular and energy resolutions were better than 0.5° and ~20 meV, respectively. The normal of the sample surface is parallel to the axis of the electron analyzer's lens, and the incidence angle of light is 60° relative to the lens axis[54]. The band dispersion image on the exit plane of the R3000 analyzer was turned around for 180° by a homogeneous magnetic field to scatter with a ferromagnetic Fe(001)-$p$(1×1)-O target and then turned around for another 180° to obtain a spin-resolved $E$-$k$ image at a fluorescence screen.

The spin-resolved $E$-$k$ images in Fig. 2e-j, Extended Data Fig. 9c(i), d(i), and 12 are multiplied by the total intensity ($I_{tot} = I_{\text{mag up}} + I_{\text{mag down}}$) for better visualization of spin polarization on the energy bands, where $I_{\text{mag up}}$ ($I_{\text{mag down}}$) is the photoelectron intensity measured under opposite ferromagnetic target magnetization, while those in Extended Data Fig. 5, 6, and 9c(ii), d(ii) are raw SARPES $E$-$k$ images where the intensity in each pixel is the unaltered result of Pol. = $(1/S_{eff}) \times (I_{\text{mag up}} - I_{\text{mag down}}) / (I_{\text{mag up}} + I_{\text{mag down}})$, where $S_{eff} = 0.32$ is the Sherman function. Values in the polarization curves in Fig. 2 were also defined as Pol. = $(1/S_{eff}) \times (I_{\text{mag up}} - I_{\text{mag down}}) / (I_{\text{mag up}} + I_{\text{mag down}})$. The intensity of spin-up (spin-down) photoelectrons ($I_{up}$/$I_{down}$) in Fig. 2 is calculated through $I_{up} = (1 + \text{Pol.}) \times I_{tot} / 2$ and $I_{down} = (1 - \text{Pol.}) \times I_{tot} / 2$.

The spin-polarized images were obtained by iterating through multiple loops (L1 – L20) within an integration time of about 14 hours. In Extended Data Fig. 13a-d, we found that the sample is slightly hole-doped and the energy bands move rigidly upward to the Fermi level during the measurement process, which results in an inaccurate spin polarization value close to $E_F$. To eliminate the artifact caused by hole-doping, an energy offset procedure was done referring to the first loop (L1) in Extended Data Fig. 13b, d. It should be emphasized that the energy offset has a negligible effect on the spin-polarization for $E_B > 0.1$ eV, by comparing the results with or without it (Extended Data Fig. 13e).

Data on multiple samples show consistent SARPES $E$-$k$ images, confirming the repeatability of our measurements (Extended Data Fig. 12).

**Synchrotron-based SARPES measurements and data analysis**

Polarization curves (in Fig. 3c) are measured at BL09U (Dreamline) of the SSRF with a Scienta-Omicron DA30 analyzer and a single VLEED spin detector. The entrance slit is perpendicular to the ground, and spin direction measured by the VLEED detector is perpendicular to the slit. Measurement temperature is 30 K. The sample was cleaved *in situ* in an ultra-high vacuum chamber with pressure better than $6 \times 10^{-11}$ mbar. The energy resolution was set to 20 meV, and the spot size of the light is $20 \times 30$ μm. Incident light is linearly polarized. The LH/LV (*p*-/*s*-) polarized light is defined as the electric vector of the light being parallel (perpendicular) to the plane formed by the incident direction of the light and the normal direction of the sample. $S_x$ polarization curves at $h\nu =$ 28, 66 and 82 eV were obtained from the same sample.

The experimental results measured using different light polarizations and photon energies can eliminate the influence of possible matrix elements[55,56]. On the one hand, the data at 21.2 eV is taken with a helium lamp which emits essentially unpolarized light. Therefore, the matrix element effect due to light polarization is eliminated. On the other hand, $S_x$ polarization curves at the "L" point at $h\nu =$ 28 eV under LH and LV polarization reveal the same trend of spin polarization: $S_x < 0$ for the most part of the curves. Moreover, since 21.2 and 28 eV light probe the same out-of-plane Brillouin zone (centered at $\Gamma_3$), whereas 66 and 82 eV light probe another zone centered at $\Gamma_5$, possible matrix element effect due to different SARPES responses on different Brillouin zones is also eliminated. Therefore, we believe that the apparent spin polarization reflects the intrinsic spin of the band.

It should be noted here that quantitative deviations to the overall trend do occur at some of the $S_x$ polarization curves in Fig. 3, but this is understandable as different bands would response slightly differently under different photon energies, and the data suffers from relatively low signal-to-noise ratio since the intrinsic spin signal is weak. Importantly, a qualitative agreement is met between the theoretical three-dimensional spin texture and the measured $S_x$ polarization.

**Scanning NV magnetometry measurements**

The scanning NV magnetometry[57-59] is applied to visualize the local magnetic field from the single-crystal $MnTe_2$. The NV center is implanted in the apex of a pillar etched from a diamond cantilever and is attached to the tuning fork of an atomic force microscope provided by Qzabre. The NV spin state is detected through the optical excitation at 532 nm with the power of ~100 μW so that the local magnetic field is extracted based on the shift in the optically detected magnetic resonance (ODMR). To efficiently apply the microwave to the NV center, we employ the copper wire with a 20 μm diameter on the top of the single-crystal $MnTe_2$ as shown in Extended Data Fig. 10a. The sample is characterized by a high spatial resolution (~30 nm) which is only limited by the separation distance between the NV center and the surface of the crystal. We have scanned over a larger scan area (50 μm) to confirm whether there are any local magnetization features presented in the surface stray field. All the magnetometry measurements were conducted at a temperature of 2 K, far below the Néel temperature of $MnTe_2$ and with a remnant bias field of 6 Gauss along the NV axis to split the spin-1 states which will not influence any magnetization domain of the $MnTe_2$. All two-dimensional magnetic field images are obtained by utilizing a pulsed ODMR scheme which can potentially reach the field sensitivity threshold of 2 $\mu T/\sqrt{Hz}$.

**DFT Calculations**

All DFT calculations with and without SOC were performed within the Perdew-Burke-Ernzernhof (PBE)[60] exchange-correlation functional using a plane-wave basis set and projected augmented-wave (PAW) method[61] as implemented in the Vienna ab initio simulation package (VASP)[62,63]. The experimental lattice constant of 6.90 Å, a Monkhorst-Pack (9×9×9) $k$-point mesh[64], and an energy cutoff of 500 eV have been used. The standard PBE pseudopotential is adopted in all calculations, treating 13 valence electrons for Mn ($3p^63d^54s^2$) and six valence electrons for Te ($5s^25p^4$). To account for the correlation effects of the Mn $3d$ electrons, the rotationally invariant Dudarev's formalism was performed with the reported $U = 5.0$ and $J = 0.8$ eV[65,66]. Atomic positions were optimized until the Hellman-Feynman force on each atom was smaller than 0.01 eV/Å and the electronic iteration was performed until the total energy change was smaller than $10^{-6}$ eV. The spin-projected band structures are analyzed using the PyProcar code[67].

To calculate the surface states, we constructed the Wannier tight-binding Hamiltonian obtained from WANNIER90 code[68-70]. In the tight-binding Hamiltonian construction, 88 Wannier functions, including the Mn-$d$ and Te-$p$ orbitals, were chosen. Finally, the surface states $E$-$k$ images, CECs, and surface spin polarization were calculated using the retarded Green's function of semi-infinite models of $MnTe_2$. The Wannier interpolation approach with 501×501 (501) crystal momentum points was adopted in WannierTools[71] for calculations of CECs ($E$-$k$ images).

**$k \cdot p$ Model**

To theoretically identify the spin texture in $MnTe_2$, we constructed the $k \cdot p$ Hamiltonian near the Γ point according to the magnetic point group (little co-group) symmetries. The little co-group of the Γ point in MnTe$_2$ bulk is *m-3*, which contains space-inversion symmetry but breaks time-reversal symmetry. The representation matrices of the generators of *m-3* with the spin basis $(|\uparrow\rangle, |\downarrow\rangle)$ were written as:

$$A^\Gamma(3_{111}^+) = e^{-i\pi/3\,(\sigma_x+\sigma_y+\sigma_z)/\sqrt{3}}$$
$$A^\Gamma(2_{100}) = -i\sigma_x$$
$$A^\Gamma(2_{001}) = -i\sigma_z$$
$$A^\Gamma(\bar{1}) = \sigma_0,$$

where $\sigma_i$ denote the Pauli matrices. Through the theory of invariants $A^\Gamma(g)H^\Gamma(\mathbf{k})A^\Gamma(g)^{-1} = H^\Gamma(g\mathbf{k})$, the Hamiltonian near the Γ point can be obtained:

$$H^\Gamma = h_1\sigma_0 + h_3(k_x^2 + k_y^2 + k_z^2)\sigma_0 + h_3(k_y k_z \sigma_x + k_z k_x \sigma_y + k_x k_y \sigma_z) + O(\mathbf{k}^3)$$

where $h_i$ are the undetermined coefficients. To directly display the in-plane spin texture, the out-of-plane spin polarization terms and spin-independent terms could be omitted:

$$H_{in}^\Gamma = h_3 k_z(k_y \sigma_x + k_x \sigma_y) + O(\mathbf{k}^3).$$

Similarly, the Hamiltonian expanded near the M point (little co-group *mmm*) has the form $H_{in}^M = h_4 k_z k_y \sigma_x + h_5 k_z k_x \sigma_y + O(\mathbf{k}^3)$. Therefore, the AFM-induced spin splitting of MnTe$_2$ bulk manifests a quadratic spin texture relative to the *k*-polynomial.

In the case of spin splitting solely induced by SOC, a non-magnetic symmetry group that preserves time-reversal symmetry should be utilized to avoid the contribution of the local moments. In this case, the bulk material exhibits both time-reversal and space-inversion symmetries, resulting in the absence of spin splitting. Therefore, spin splitting induced by SOC can only arise from the inversion symmetry breaking at the surface. The corresponding surface little co-group in MnTe$_2$ is *m1'*, and the corresponding representation matrices are expressed as:

$$A^\Gamma(m_{100}) = -i\sigma_x,$$
$$A^\Gamma(T) = -i\sigma_y \kappa,$$

where $\kappa$ is conjugate operator. The corresponding Hamiltonian is written as:

$$H^\Gamma = h_1\sigma_0 + h_2(k_y\sigma_x - k_x\sigma_y) + h_3(k_y\sigma_x + k_x\sigma_y) + h_4 k_x \sigma_z + h_5 k_x^2 \sigma_0 + h_6 k_y^2 \sigma_0 + O(\mathbf{k}^3).$$

By retaining the in-plane spin polarization terms, the Hamiltonian now reads:

$$H_{in}^{\Gamma} = h_2(k_y\sigma_x - k_x\sigma_y) + h_3(k_y\sigma_x + k_x\sigma_y) + O(\boldsymbol{k}^3),$$

which includes only linear Rashba and Dresselhaus terms.

## Method References

**Acknowledgements**

We thank Koji Miyamoto, Kenta Kuroda, Taichi Okuda, Mingzhu Zhang and Liwei Deng for the help in SARPES measurements, Ruoming Peng for the help in scanning NV magnetometry measurements, and Junxue Li for discussions. Work at SUSTech was supported by the National Key R&D Program of China (No. 2022YFA1403700 and No. 2020YFA0308900), the National Natural Science Foundation of China (NSFC) (No. 12074161 and No. 12274194), the Key-Area Research and Development Program of Guangdong Province (2019B010931001), the Guangdong Provincial Key Laboratory for Computational Science and Material Design (No. 2019B030301001), the Guangdong Innovative and Entrepreneurial Research Team Program (No. 2016ZT06D348) and Shenzhen Science and Technology Program (Grant No. RCJC20221008092722009). The DFT calculations were performed at Center for Computational Science and Engineering of SUSTech. Work at SIMIT was supported by the NSFC (No. U1632266, No. 11927807 and No. U2032207), and the National Key R&D Program of China (No. 2022YFB3608000). Work at Westlake was supported by the NSFC (No. 12274353), National Key R&D Program of China (No. 2022YFA1402200), and the Westlake Instrumentation and Service Center for Physical Sciences. D. S. acknowledges support from the NSFC (No. U2032208). C. L. acknowledges support from the Highlight Project (No. PHYS-HL-2020-1) of the College of Science, SUSTech.

**Author Contributions**

C.L. and Q.L. conceived and designed the research project. Y.-P.Z. grew and characterized the single crystals. S.Q., H.Z., and W.L. designed and built the image-type SARPES setup. Y.-P.Z., X.-R.L., H.Z., G.Q., C.H., Z.J., X.-M.M., Y.-J.H., M.-Y.Z., W.L., M.Z., J.D., S.M., K.T., M.A., Z.L., M.Y., D.S., Y.H., R.-H.H., S.Q., and C.L. performed the ARPES measurements. X.C., Y.L., P.L., J.L., and Q.L. performed the theoretical analysis and DFT calculations. S.J., M.L., and J.W. performed the scanning NV magnetometry measurements. Y.-P.Z., X.C., Q.L., and C.L. wrote the paper with the help from all authors.

**Data Availability**

The data that support the findings of this study are available from the corresponding authors upon request. Correspondence and requests of ARPES and DFT data are addressed to C.L. and the NV magnetometry data are addressed to J.W.

**Competing Interests**

The authors declare no competing interests.

**Extended Data Table 1 | The differences in the spin splitting originating from SOC and AFM magnetic order in MnTe$_2$ encompass variations in symmetry, $k_z$-dependence, temperature-dependence, $k \cdot p$ models, and permissible spin texture types.**

| Origin | SOC (surface) | | AFM (bulk) | |
|---|---|---|---|---|
| Symmetry group | *Pc1'* | | *Pa-3* | |
| Time-reversal | ✓ | | ✗ | |
| Space-inversion | ✗ | | ✓ | |
| $k_z$-dependent | ✗ | | ✓ | |
| $T$-dependent | ✗ | | ✓ | |
| Allowed spin splitting $k \cdot p$ terms | Odd only | | Even only | |
| $k \cdot p$ model | $k_y\sigma_x - k_x\sigma_y$ | $k_y\sigma_x + k_x\sigma_y$ | $k_z(k_y\sigma_x + k_x\sigma_y)$ | |
| Spin texture type | Rashba (Linear) | Dresselhaus (Linear) | Plaid-like (Quadratic) | |
| | | | ($k_z > 0$) | ($k_z < 0$) |
| Spin texture in $k_x$-$k_y$ plane | 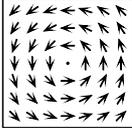 | 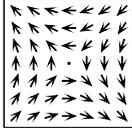 | 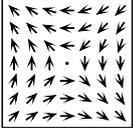 | 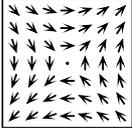 |

**Extended Data Table 2 | Little group and the allowable spin-splitting of bands at different $k$ positions.** Splitting of all spin components $S_x$, $S_y$ and $S_z$ are allowed in general points. Little groups are orders with respect to the three axes, e.g. 2mm at the Z point represents $C_{2x}$, $M_y$ and $M_z$.

| Position | High symmetry line | Little group | Spin splitting |
|---|---|---|---|
| Γ (0, 0, 0) | Not a line | m-3 | no |
| X (0, 1/2, 0) | Not a line | mmm | no |
| M (1/2, 1/2, 0) | Not a line | mmm | no |
| R (1/2, 1/2, 1/2) | Not a line | m-3 | no |
| X1 (1/2, 0, 0) | Not a line | mmm | no |
| Δ (0, v, 0) | Γ - X | m2m | no |
| Z (u, 1/2, 0) | X - M | 2mm | no |
| Σ (u, u, 0) | M - Γ | ..m | $S_z$ |
| Λ (u, u, u) | Γ - R | 3 | $S_x$, $S_y$, $S_z$ |
| S (u, 1/2, u) | R - X | .m. | $S_y$ |
| T (1/2, 1/2, w) | R - M | mm2 | no |
| ZA (1/2, u, 0) | M - X1 | m2m | no |
| Δ' (u, 0, 0) | X1 - Γ | 2mm | no |
| P1 (u, v, 0) | Γ - X - M plane | ..m | $S_z$ |
| O (0, 0, 1/5) | Not a line | mm2 | no |
| P2 (u, 0, 1/5) | O - A | .m. | $S_y$ |
| A (1/2, 0, 1/5) | Not a line | mm2 | no |
| P3 (1/2, v, 1/5) | A - B | m.. | $S_x$ |
| B (1/2, 1/2, 1/5) | Not a line | mm2 | no |
| P4 (u, u, 1/5) | B - O | 1 | $S_x$, $S_y$, $S_z$ |
| GP (u, v, w) | General points | 1 | $S_x$, $S_y$, $S_z$ |

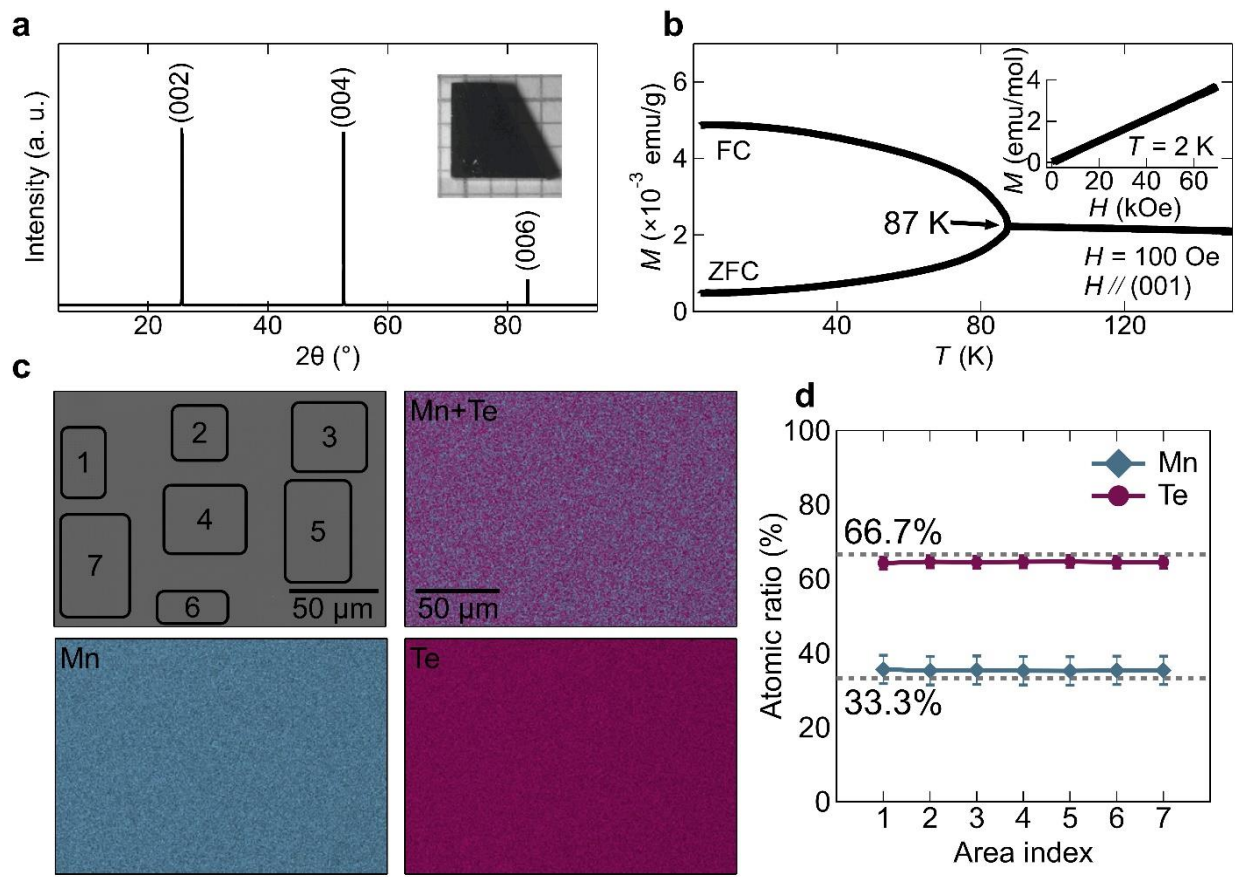

**Extended Data Fig. 1 | Structural, magnetic, and compositional characterization of MnTe₂ single crystals. a**, Single crystal x-ray diffraction results of MnTe$_2$, showing a (001) exposed plane. Inset: a MnTe$_2$ single crystal against a millimeter grid. **b**, Field-cooled (FC) and zero-field-cooled (ZFC) temperature dependence of magnetization with $H$∥(001). The magnetic transition temperature is $T_N$ = 87 K. The downturn of the ZFC $M$-$T$ curve at $T_N$ signals the antiferromagnetic behavior. Inset: magnetic field dependence of magnetization at $T$ = 2 K. **c**, A typical scanning electron microscopy image of the MnTe$_2$ single crystal, and corresponding energy-dispersive EDS elemental maps, indicating a homogeneous distribution of the Mn and Te elements. **d**, Atomic ratio of Mn and Te at the seven regions indexed in **c**. A ratio close to 1:2 is found in all regions, indicating a high quality of the crystals.

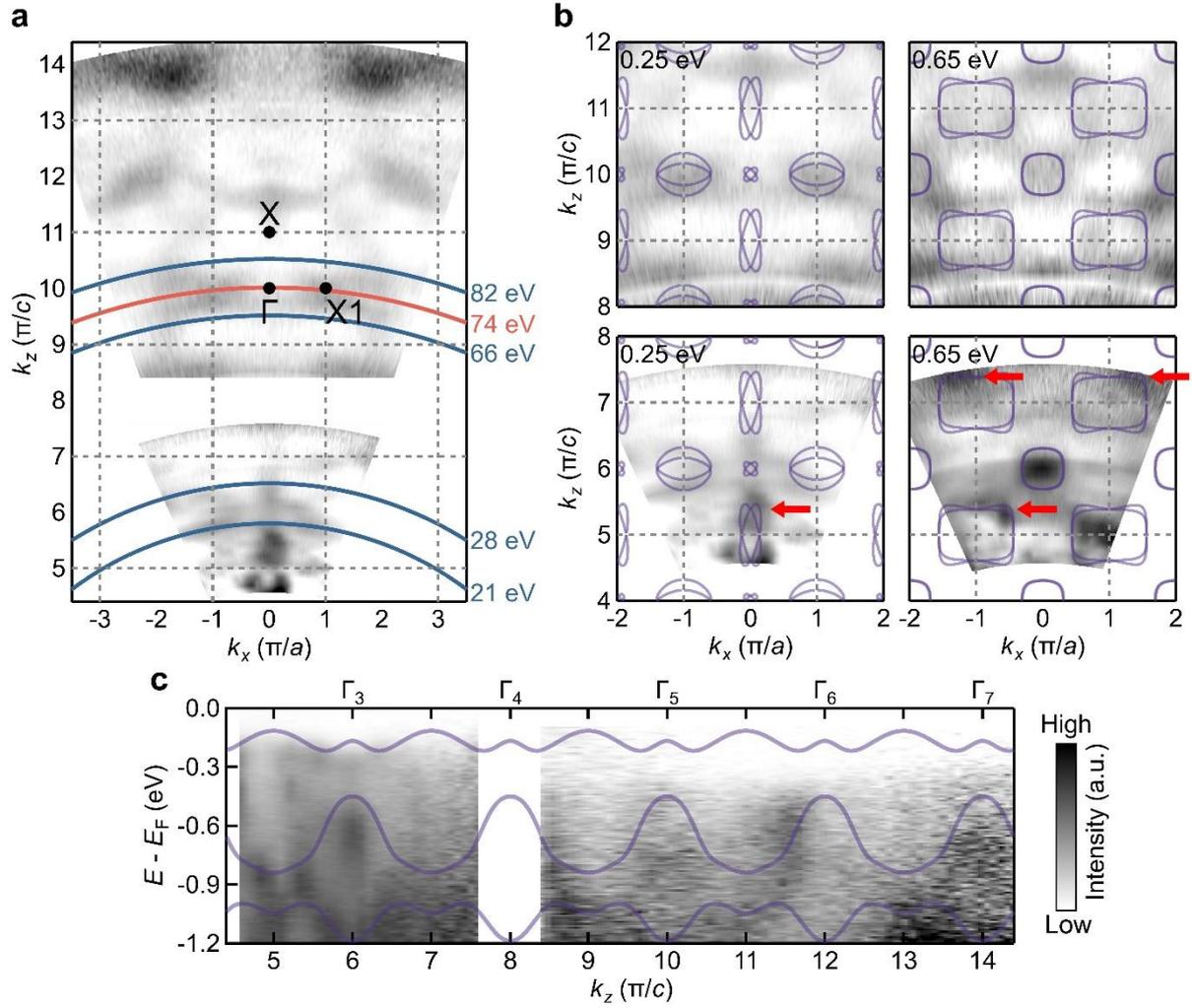

**Extended Data Fig. 2 | Electronic structure along the $k_x$-$k_z$ plane. a**, ARPES $k_x$-$k_z$ map at binding energy $E_B = 0.25$ eV. A clear $k_z$ dispersion is seen, consistent with the bulk nature of the observed bands. The inner potential is determined to be $V_0 = 10$ eV from the data; high symmetry points are marked accordingly. The red curve in (a) shows the $k_x$-$k_z$ position of an ARPES measurement using 74 eV incident light, corresponding to $k_z = 10\ \pi/c$ ($\Gamma_5$) at the in-plane Brillouin zone center. The cyan curves mark the ARPES measurement positions in Fig. 3 of the main text. $h\nu = 21.2$, 28, 66, and 82 eV corresponds to $k_z = 5.8$ (−0.2), 6.5 (0.5), 9.5 (−0.5) and 10.5 (0.5) $\pi/c$, respectively. **b**, $k_x$-$k_z$ dispersion at binding energies $E_B = 0.25$ and 0.65 eV, with corresponding CECs calculated by DFT. The experimental spectral intensities show rough agreement with the calculated results (indicated by red arrows). **c**, $E$-$k_z$ dispersion along $k_x = 0$. The band at $E_B = 0.5$–1.0 eV is seen to experience repetitive dispersion within our measurement range, which matches qualitatively with that of a theoretical bulk band in the dispersing period, bandwidth, and energy locations of band tops.

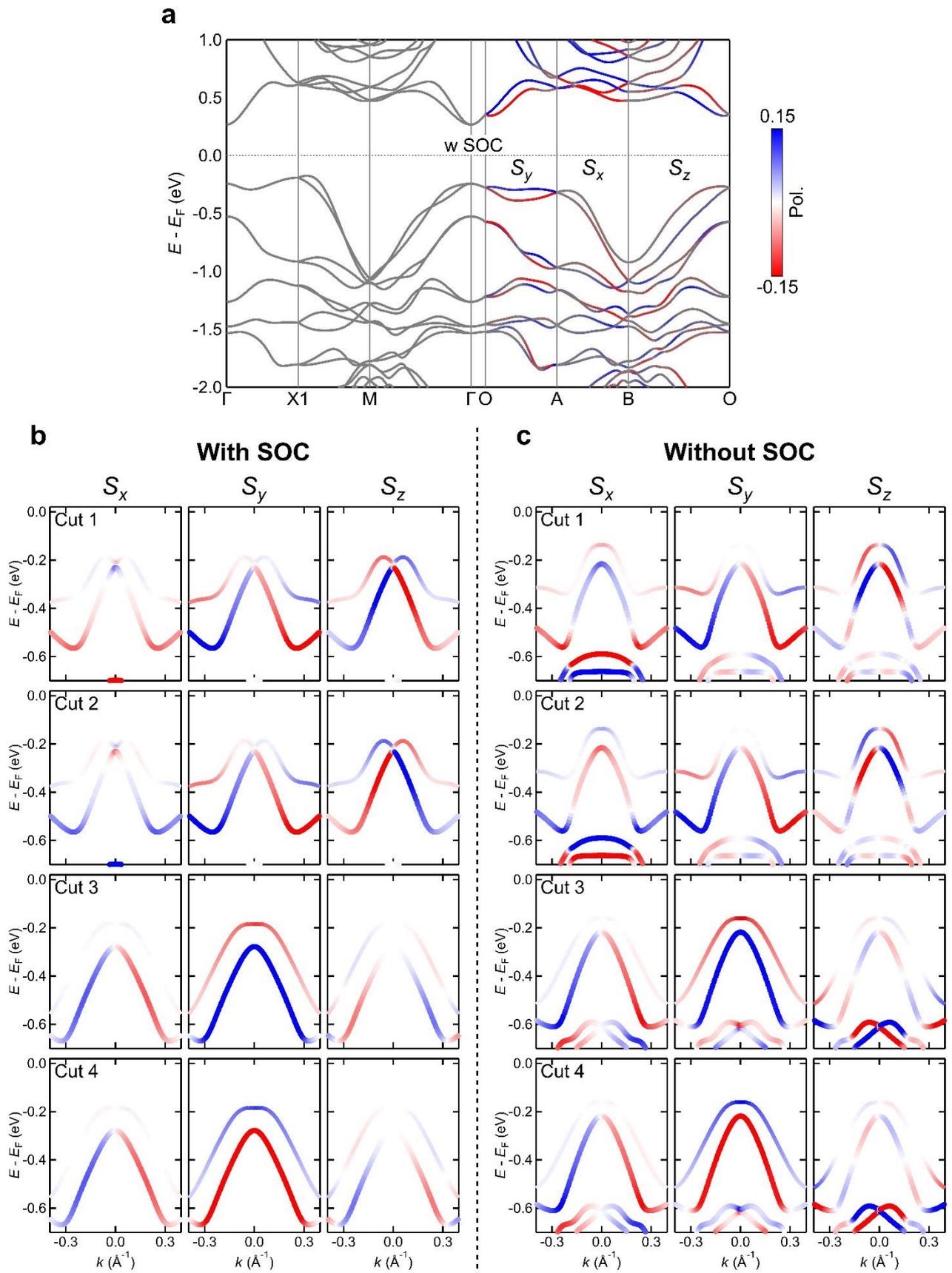

**Extended Data Fig. 3 | DFT-calculated spin-resolved energy bands with and without SOC. a**, *E-k* dispersion with SOC. A high level of spin polarization also exists in the O-A-B-C plane when SOC is turned on. A comparison between this result and Fig. 1j proves that the existence of SOC is not a necessary condition for the AFM-induced spin splitting. Even though Te is a heavy element that has strong SOC, the relativistic spin-orbit interaction gives only mild, secondary effects on the spin splitting of the $MnTe_2$ bulk bands. **b,c**, DFT-derived spin-resolved *E-k* dispersion along Cuts 1 – 4 with and without SOC, respectively. All calculated results show a plaid-like antisymmetric spin texture.

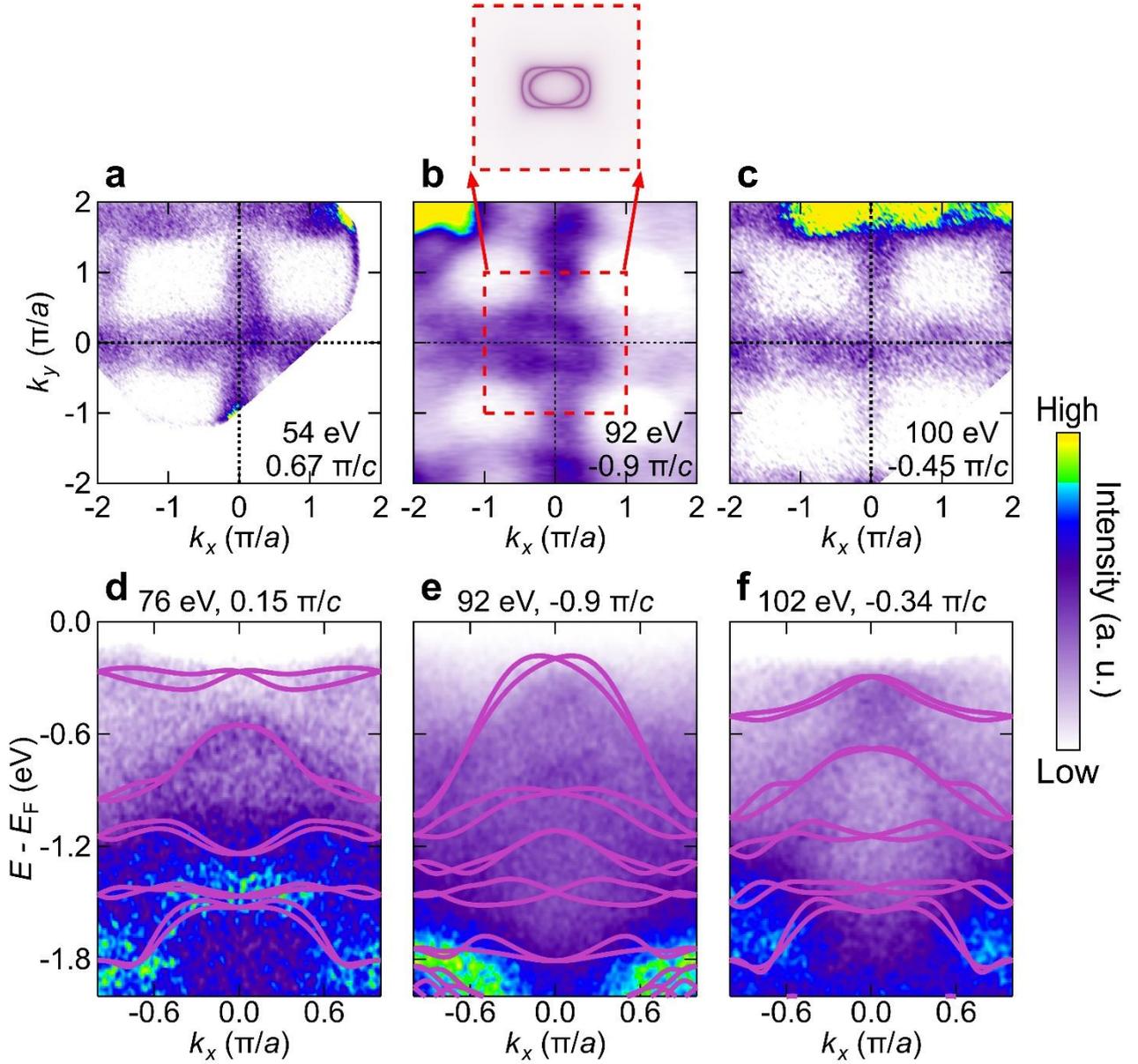

**Extended Data Fig. 4 | Electronic structure at different photon energies. a-c**, CECs at binding energy $E_B = 0.35$ eV measured with 54, 92, and 100 eV photons. The CEC at 92 eV show a rectangular feature close to the zone center (in good agreement with the DFT result shown in the inset), whereas the other two maps show a cross shape in the first zone. This $k_z$ dispersive behavior indicates the bulk nature of the bands. **d-f**, ARPES $E$-$k$ dispersion taken with 76, 92, and 102 eV photons. Purple curves are the bulk bands calculated by DFT. The qualitative match between the DFT bands and the ARPES data likely indicates that the ARPES intensity comes mostly from bulk states of the system.

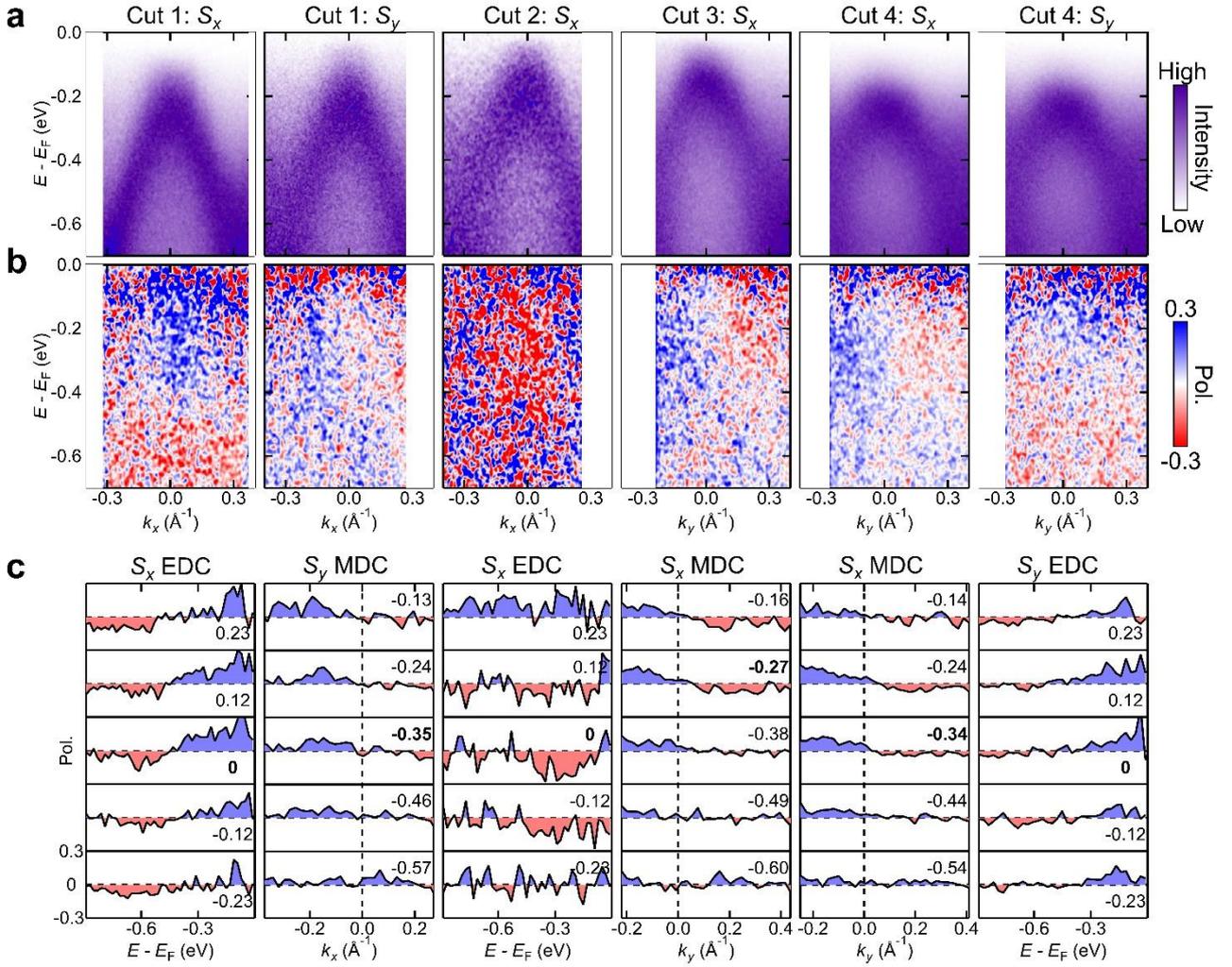

**Extended Data Fig. 5 | Raw spin-resolved *E-k* images along Cuts 1 – 4 and corresponding polarization curves. a**, Spin-integrated *E-k* images along Cuts 1 – 4 obtained by the spin-imaging ARPES system, shown in Fig. 2e-j. **b**, Corresponding raw spin-resolved *E-k* images. The difference between these images and those in Fig. 2 is that the data here is not multiplied by the spin-integrated intensity of the bands. In other words, these are the "raw" spin-polarized data obtained with the image-type spin detector. **c**, Spin polarization curves of Cuts 1 – 4 at all measured *E-k* areas. From these images and polarization curves, the same conclusions as those in the main text can be drawn: the *x* (*y*) component of the spin polarization vector is antisymmetric about $k_y = 0$ ($k_x = 0$). Additionally, we see that non-vanishing spin polarization also exists in *E-k* areas that have "no bands" (low spin-integrated ARPES intensity), which possibly comes from spin signals of vague electronic states or from bands at other $k_z$'s ($k_z$ broadening effect).

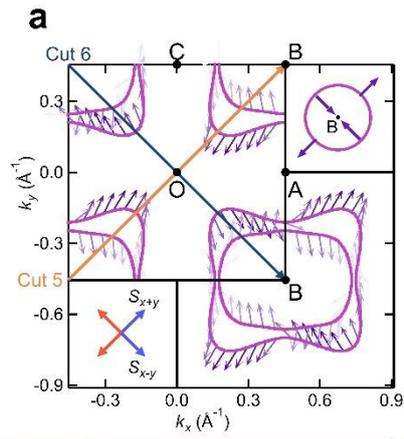
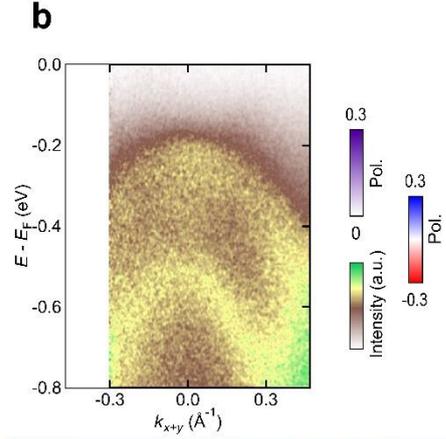
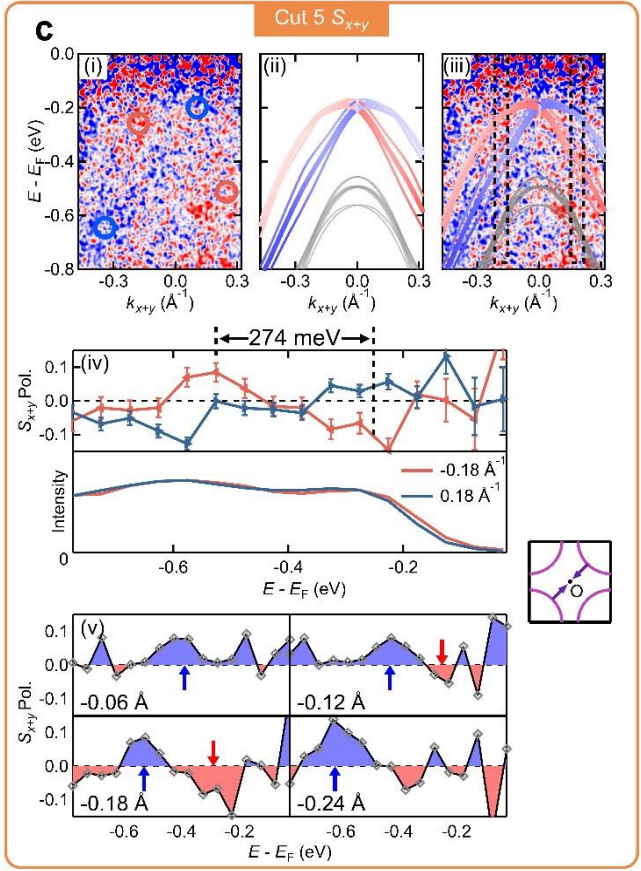
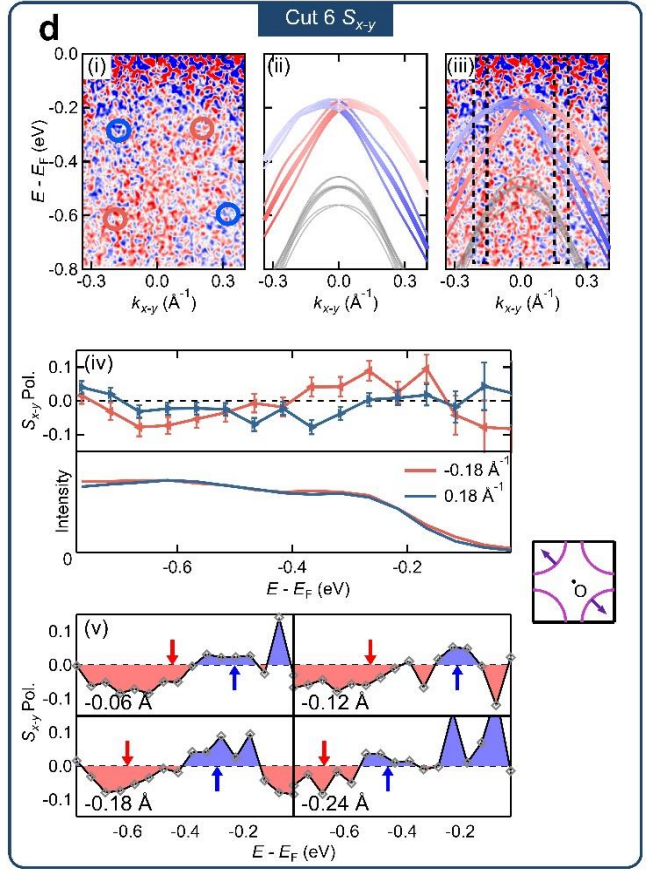

**Extended Data Fig. 6 | Additional evidence for the Dresselhaus-like in-plane spin texture. a**, DFT-calculated spin-resolved CEC at $k_z = -0.2\ \pi/c$ and binding energy $E_B = 0.45$ eV. In-plane spins point mainly along $\pm S_{x+y}$ and $\pm S_{x-y}$, resembling the Dresselhaus configuration. The direction of $+S_{x+y}$ and $+S_{x-y}$ are defined along $k_x + k_y$ and $k_x - k_y$ respectively, shown in the bottom left inset. The top right inset depicts the schematic spin texture centered at point B. **b**, Spin-integrated ARPES band dispersion along Cut 5. Cuts 5 – 6 are defined in **a**; the bands along Cut 6 (not shown) appears the same. The observed bands exhibit two hole-pockets and additional intensity at $k = 0$ and $E_B = 0.4$ eV. **c,d**, $S_{x+y}$ and $S_{x-y}$-resolved raw $E$-$k$ cuts along Cuts 5 – 6. Blue/red circles highlight the regions with +/– polarizations. The spin-resolved ARPES results (i) vaguely corroborate with the DFT-calculated results (ii), which becomes more obvious by overlaying the calculated spin-resolved energy bands on the spin-resolved ARPES images (iii). (iv) Polarization curves (top) and spin-integrated intensity curves (bottom) at $k = \pm 0.18$ Å$^{-1}$, integrated within the dashed rectangles in (iii). Each point on a curve is an integrated intensity over a $(E, k)$ range of (50 meV, 0.08 Å$^{-1}$). These spin-resolved EDCs reveal three important features. First, the $S_{x+y}$ polarization at $k_{x+y} = \pm 0.18$ Å$^{-1}$ have opposite polarization at the same binding energy. Second, the polarization at low binding energy (about 0.25 eV) and high binding energy (about 0.55 eV) is opposite. The corresponding $S_{x-y}$ polarization of Cut 6 also exhibit these two features. Third, comparing the polarization curves at –0.18 Å$^{-1}$ in Cut 5 and Cut 6, one finds that their polarizations are opposite at the same binding energy. Extracted from the polarization curves, the energy scale of the spin splitting is about 274 ± 40 meV, which is in good agreement with the calculated 297 meV and is comparable to the well-known giant bulk Rashba effect[37]. (v) Polarization curves v.s. momentum. One can trace the splitting along $k$ and find that the $+S_{x+y}$ ($-S_{x-y}$) spin-polarized peaks of Cut 5 (Cut 6) move downward in energy when $k$ moves away from O. In summary, our SARPES data along Cuts 5 and 6 reveals a Dresselhaus-like in-plane spin texture and ruled out a Rashba-like spin texture.

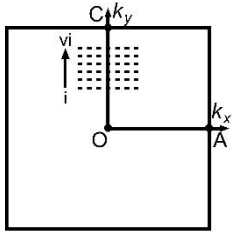

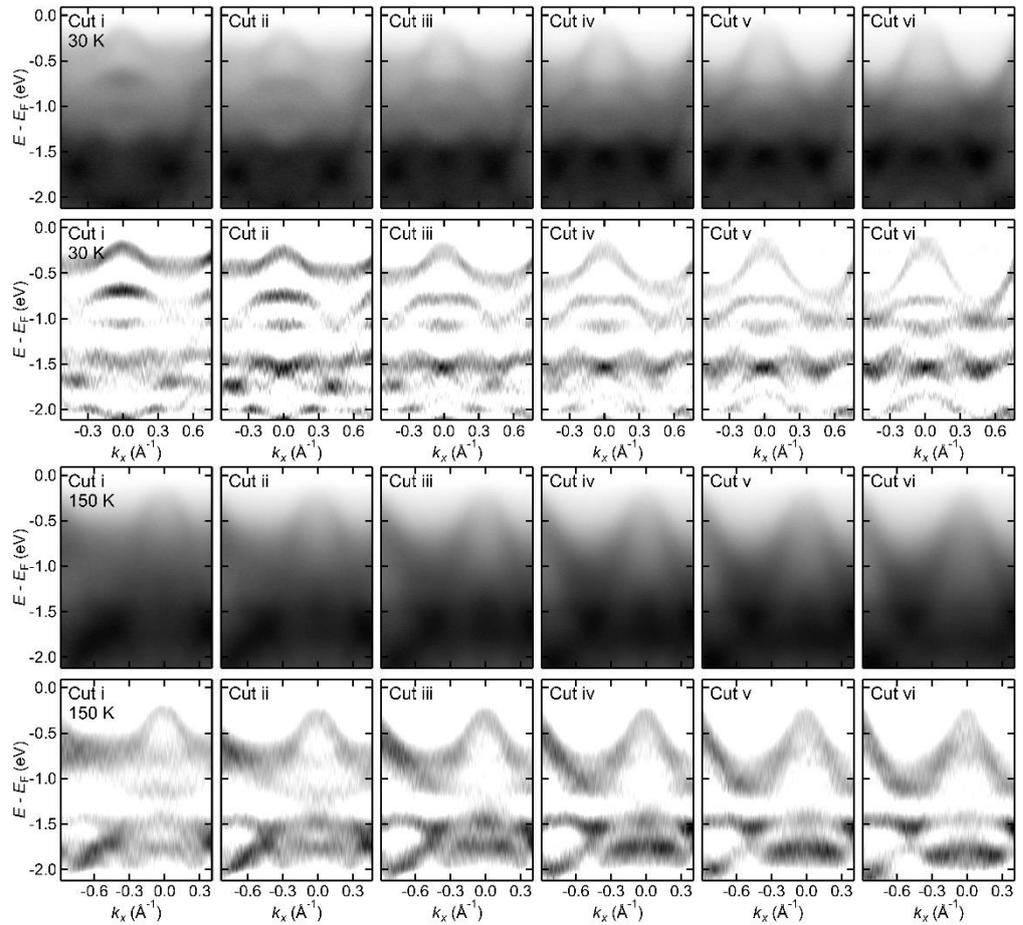

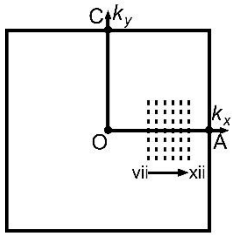

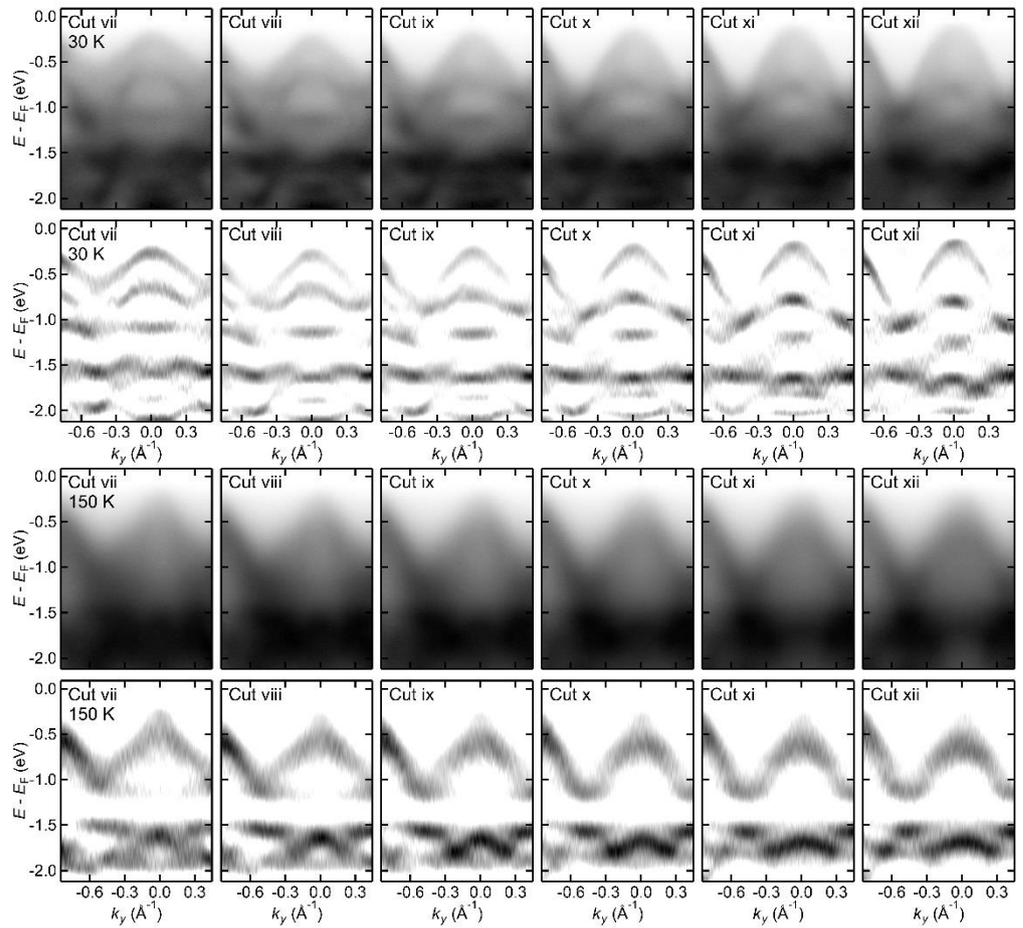

**Extended Data Fig. 7 | Temperature dependence of the ARPES band structures.** ARPES band dispersion along Cuts i – xii and corresponding second derivative analysis along the energy direction at $T = 30$ K and 150 K, respectively. The positions of Cuts i – xii are marked in the schematic of the first Brillouin zone. Besides the data shown in Fig. 4, here the band structure is also seen to undergo a non-rigid shift and the number of energy band changes across the Néel temperature ($T_N = 87$ K).

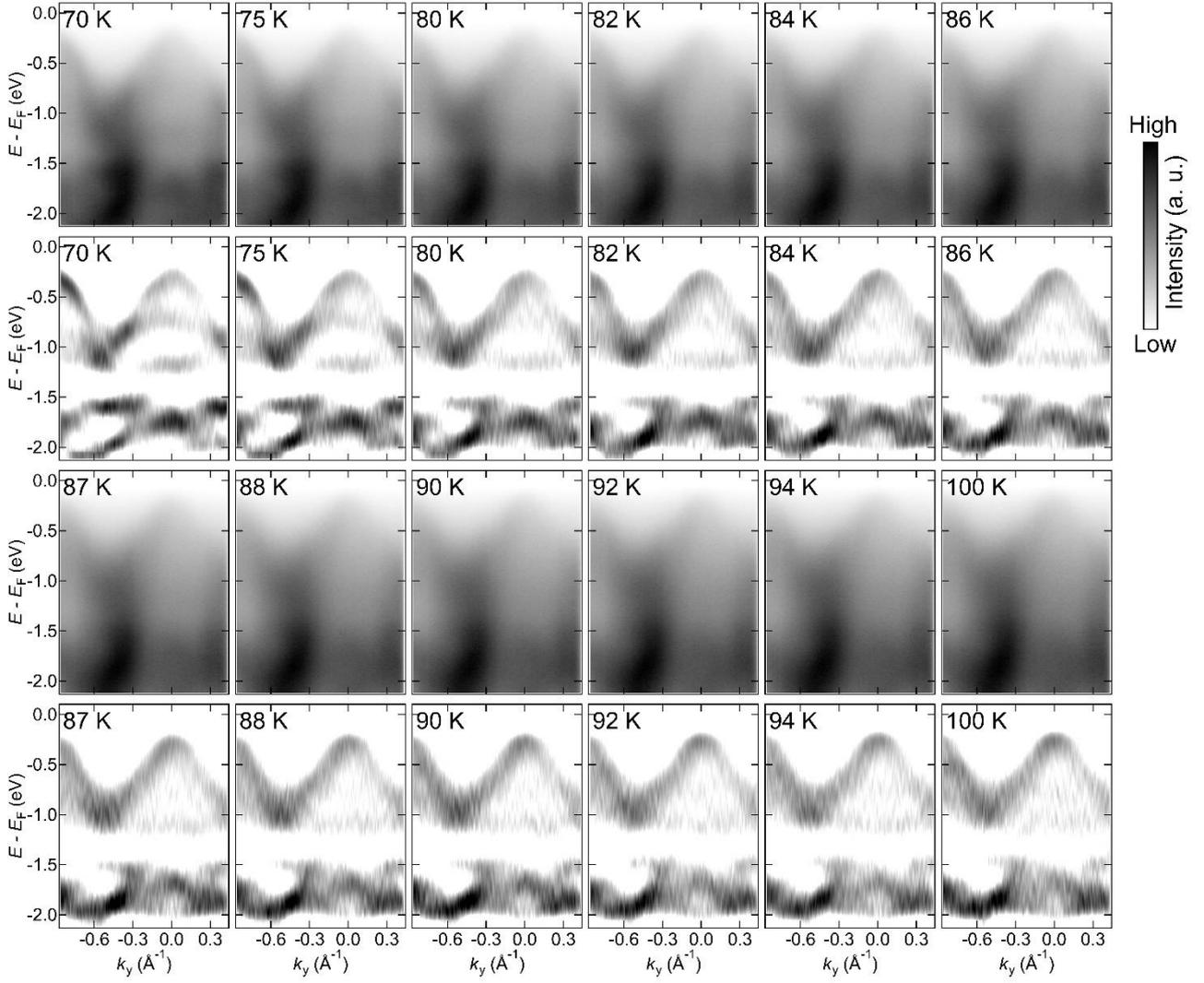

**Extended Data Fig. 8 | Temperature dependent electronic structure along Cut 3.** ARPES band dispersion below and above $T_N = 87$ K, and corresponding second derivative analysis along the energy direction. The band structures are seen to undergo a structural modification associated with the antiferromagnetic to paramagnetic transition.

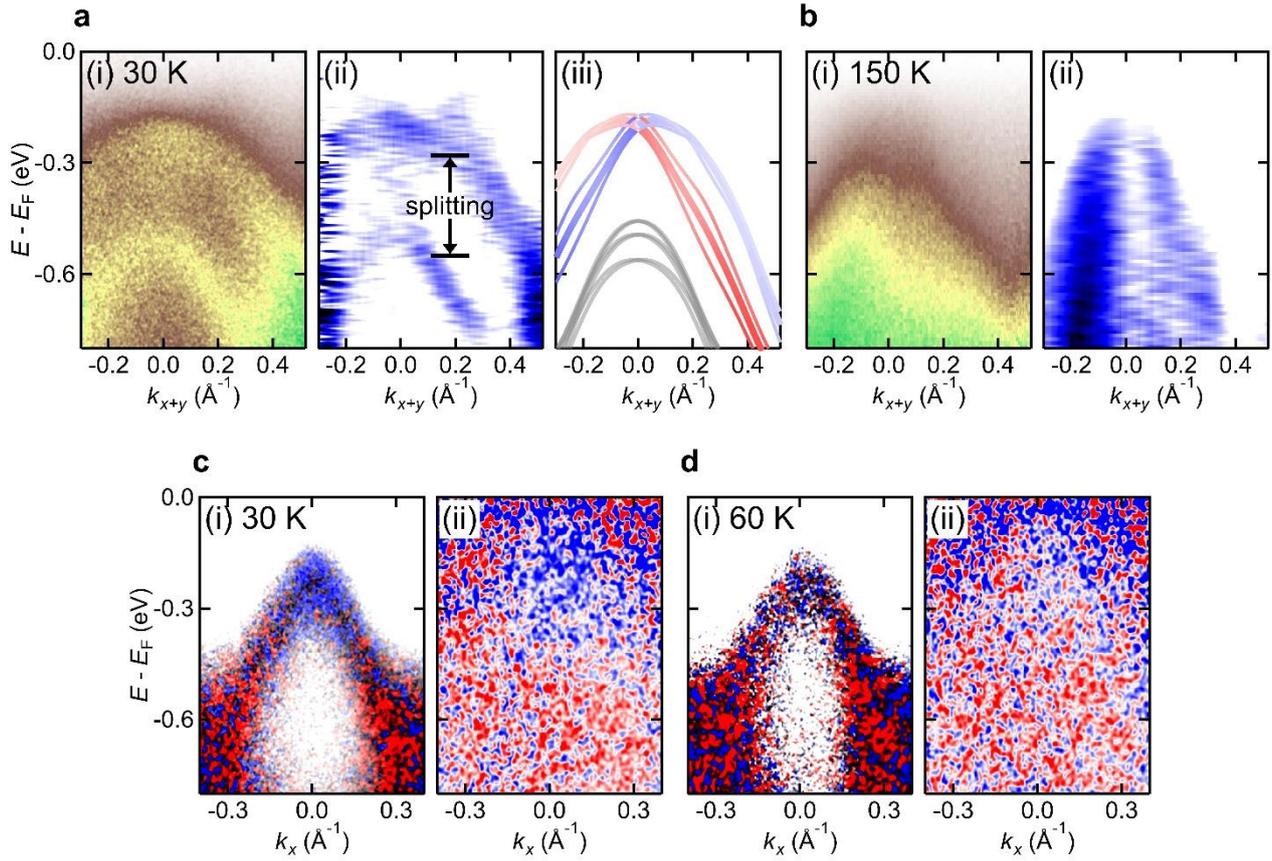

**Extended Data Fig. 9 | Temperature evolution of the spin splitting and the spin polarization signal. a,b**, Temperature dependence of the band splitting: (i) ARPES band dispersion along Cut 5 at 30 K and 150 K, (ii) corresponding second derivatives to the raw data, and (iii) spin-polarized DFT bulk bands for the low-$T$ AFM state. The abrupt change of bands between the two temperatures is evident for the vanishing of spin splitting above $T_N$. **c,d**, $S_x$-resolved $E$-$k$ images along Cut 1 and corresponding raw SARPES data. Data is taken at 30 K (**c**) and 60 K (**d**). Data in **c** and **d** are measured consecutively on the same sample with the same experimental settings and integration time. Since the bands at 30 and 60 K, as well as the associated matrix elements, are not expected to change as they belong to the same magnetic phase, the drop of spin polarization here is likely intrinsic, related to the origin of spin splitting.

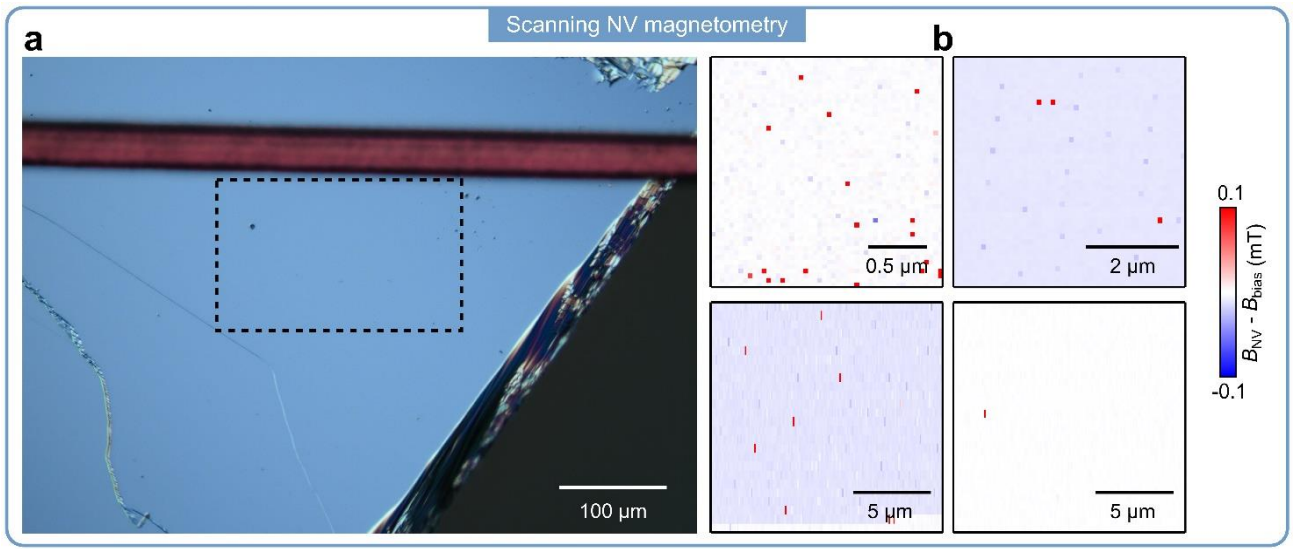
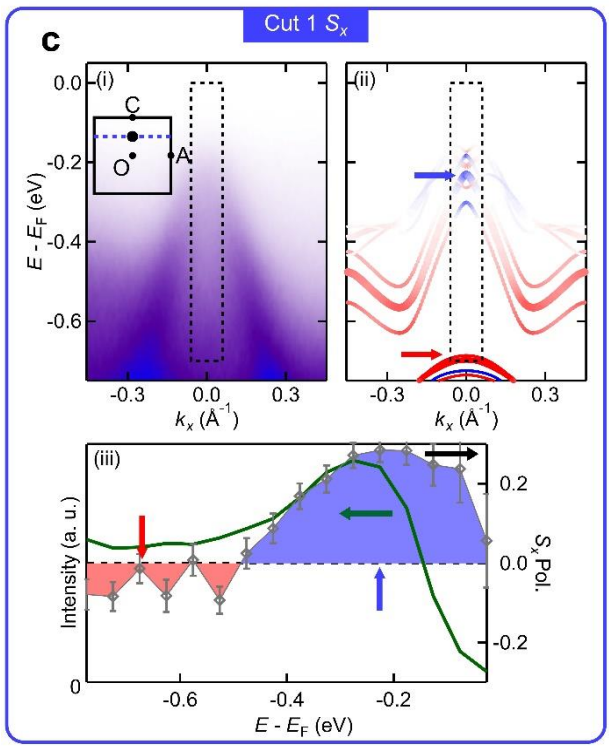
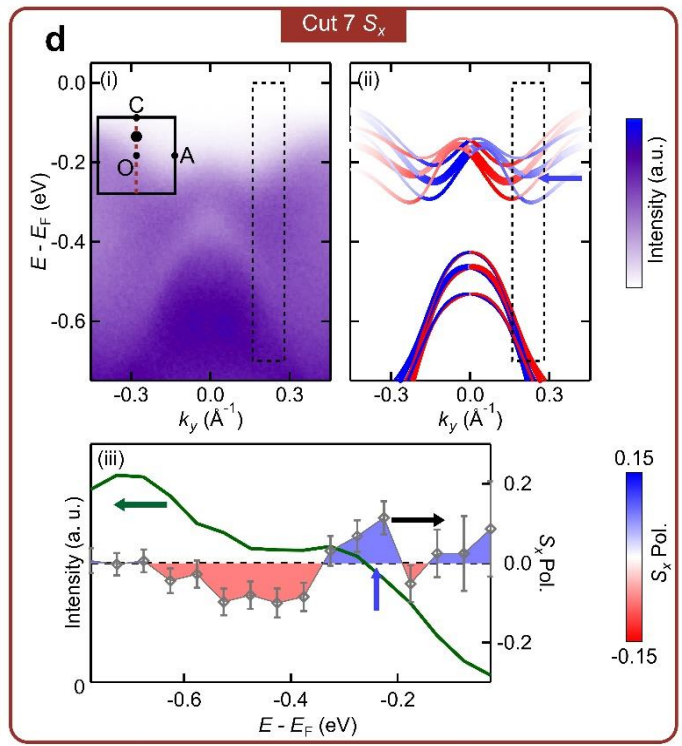

**Extended Data Fig. 10 | Evidence for a large magnetic domain size. a,b**, Evidence from cryogenic scanning NV magnetometry. **a**, A typical micrograph of the MnTe$_2$ sample. Four different areas, corresponding to **b**, are measured within the dashed rectangle. **b**, Stray magnetic fields along the NV center axis (about 55° to the surface normal) measured at 2 K. The sample is scanned using a pulsed ODMR scheme (see Methods). The spatial resolution is set to ~30 nm. An external magnetic field of 0.6 mT is applied along the NV center axis to lift the degeneracy of the NV ±1 spin states. No feature was detected within the measured length scales over a larger scan area (50 µm), indicating that the stray field variations of the sample fall below the sensitivity threshold of the NV probe ($2\ \mu T/\sqrt{Hz}$). Since the NV spin would detect non-vanishing magnetic fields contributed by domain walls, the measured feature in MnTe$_2$ suggests a uniform and large magnetic domain with no net magnetic moment contributions[57,58]. **c,d**, Evidence from $S_x$-resolved electronic structures before and after rotating the sample by 90°. (i) Spin-integrated *E-k* dispersion along Cut 1 and Cut 7 (defined in the insets). (ii) Corresponding DFT-calculated $S_x$-resolved *E-k* dispersion. (iii) Spectral intensity and $S_x$ polarization at $(k_x, k_y) = (0, 0.5\ \pi/a)$ (large black dot in the insets) measured along Cut 1 and Cut 7. These curves are integrated within the dashed rectangles in (i). The definition of $S_x$ and error bar are the same as Fig. 2. The $S_x$ polarization curves have the same sign at similar binding energies before and after rotating the sample by 90°, which is reproduced by DFT calculations (blue/red arrows). This is indicative of a magnetic domain size comparable to the size of the incident beam spot.

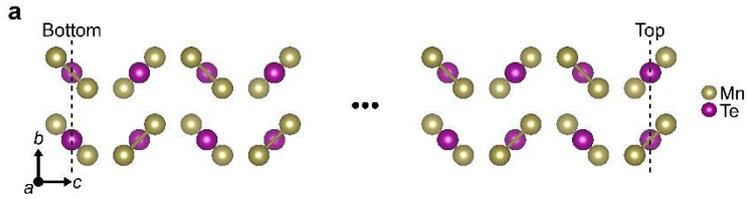

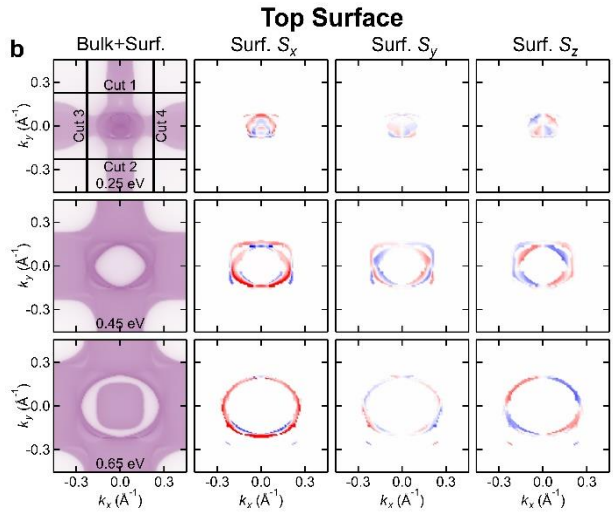

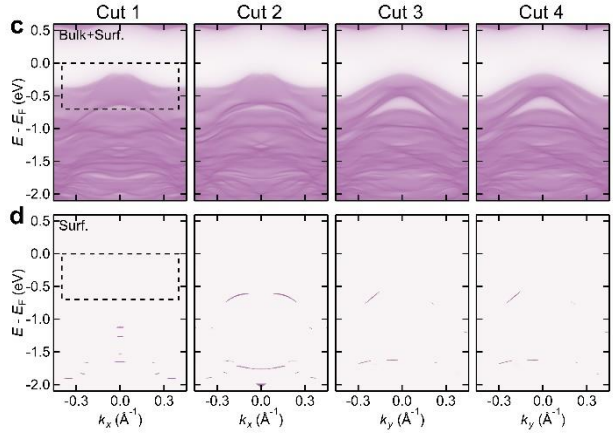

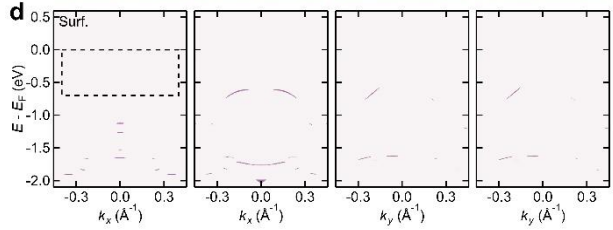

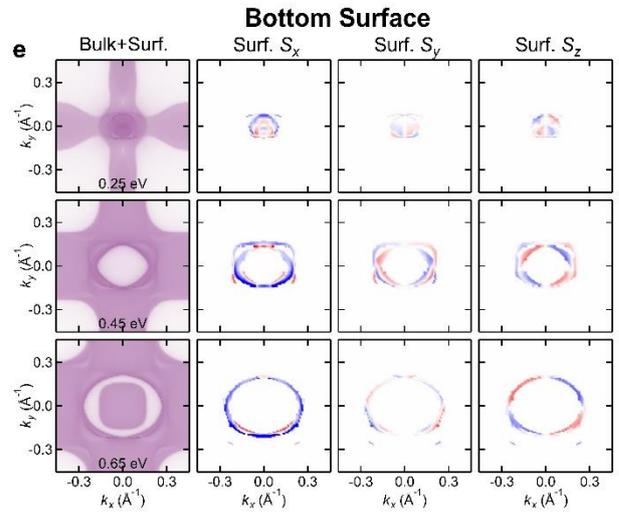

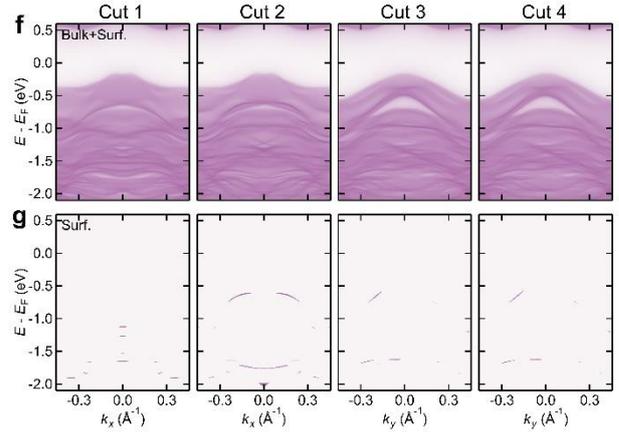

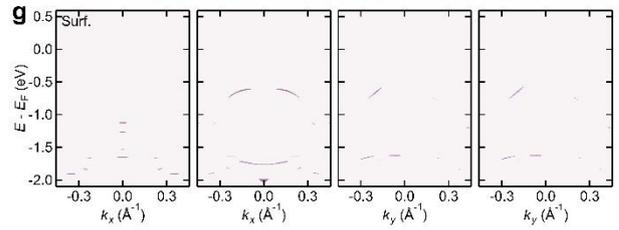

**Extended Data Fig. 11 | DFT-calculated electronic structures from semi-infinite slab models. a**, Calculation model with two different surfaces marked as "top" and "bottom". Panels **b**-**d** / **e**-**g** show the calculated results on the top / bottom surface. **b**, Spin-integrated and spin-resolved CECs at binding energies $E_B$ = 0.25, 0.45, and 0.65 eV, respectively. The spin-integrated CECs contain the spectral weights of both the bulk and the surface states. The spin-resolved CECs show the spectral weights of surface states only. At the three binding energies, there is almost no spin-polarized surface states at the positions of Cuts 1 – 4. **c**, Spin-integrated $E$-$k$ dispersions along Cuts 1 – 4. These images contain the spectral weights of both the bulk and the surface states. **d**, Corresponding spin-integrated $E$-$k$ dispersion of the surface states only. The dashed rectangles in first panel of **c** and **d** mark the ARPES-measured $E$-$k$ region along Cuts 1 – 4 in Fig. 2. Clearly, there is almost no surface state in the areas we have measured. **e**-**g**, Results of the bottom surface, yielding the same conclusion. The data in this figure is evident for the bulk nature of the spin-polarized bands in Fig. 2.

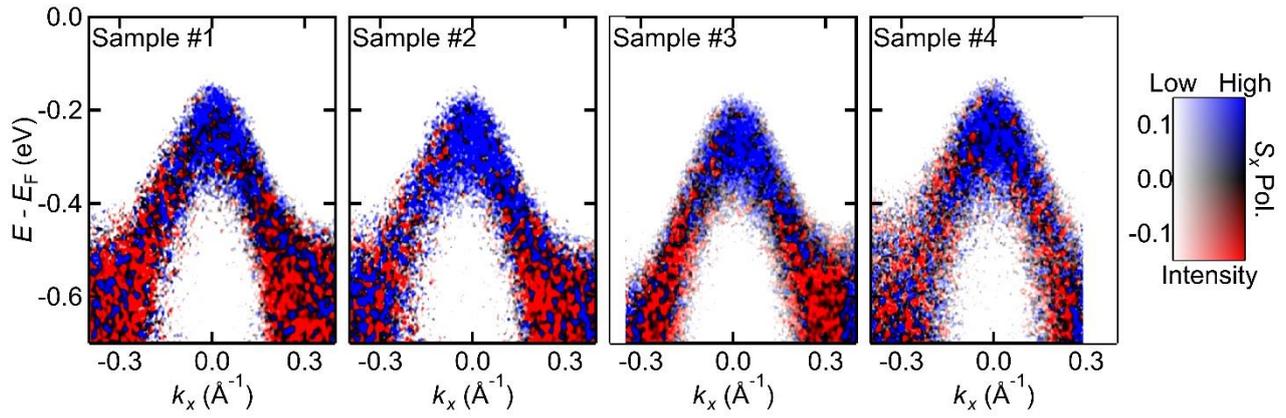

**Extended Data Fig. 12 | Repeatability of the SARPES data.** $S_x$-resolved ARPES maps on Cut 1 (Fig. 2e) are measured with four different samples. The experimental results are clearly reproducible.

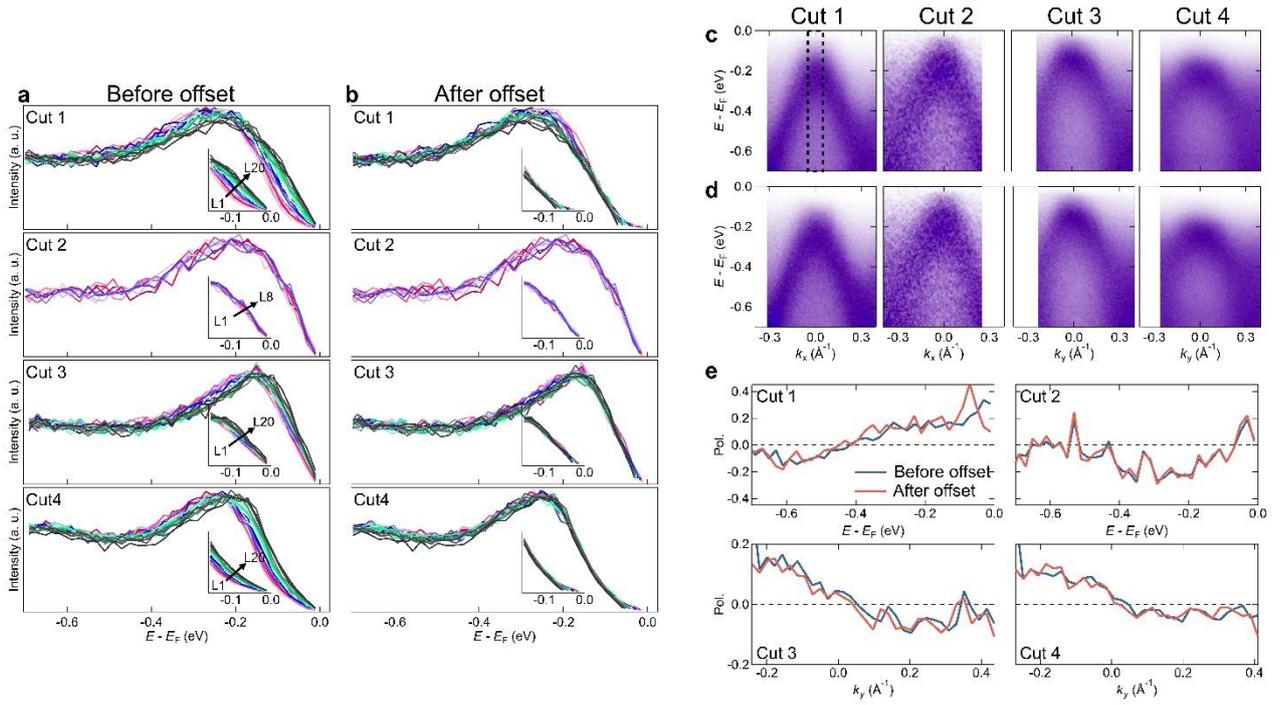

**Extended Data Fig. 13 | EDCs at $k = 0$, band dispersions, and $S_x$-polarization curves before and after an offset along the energy direction.** This process eliminates the slight hole doping effect caused by the slow but gradual aging of the samples in the vacuum chamber. **a**, EDCs of Cuts 1 – 4 before the offset. The energy bands are found to shift rigidly upward along the energy direction with increasing loop number (measurement time). **b**, EDCs of Cuts 1 – 4 after the offset. The bands are offsetting downward referring to the first loop (L1) which represents the pristine energy positions of the bands. The EDCs are taken at $k = 0$ Å$^{-1}$, integrated within the rectangles in the first panel of **c** and **d**. **c,d**, Band dispersions of Cuts 1 – 4 before and after the offset, respectively. **e**, $S_x$-polarization curves before and after the offset. The integrated positions are the same as Fig. 2e-h. One can find that the offsetting procedure introduced no qualitative difference to the spin polarization, except for a slight change of polarization magnitude near the Fermi level. The main conclusion of the paper is valid regardless of using the data before or after the offset.